\documentclass[amstex,epsf,amssymb,usenatbib]{mn2e}

\usepackage{epsfig}
\usepackage{graphicx}
\usepackage{tabularx}
\usepackage{amsmath}
\usepackage{amssymb}
\usepackage{amstext}
\usepackage{amsfonts}

\def\aj{AJ}                   
\def\araa{ARA\&A}             
\def\apj{ApJ}                 
\def\apjl{ApJ}                
\def\apjs{ApJS}               
\def\aap{A\&A}                
\def\mnras{MNRAS}             
\def\nat{Nature}              

\title{Evolutionary tracks of massive stars during formation}

\author[M.D. Smith, {\it et al.}]
{Michael D. Smith $^{1}$\thanks{E-mail: m.d.smith@kent.ac.uk} \\
$^1$Centre for Astrophysics \& Planetary Science, University of Kent, Canterbury CT2 7NH, UK
}

\date{Accepted .....
      Received ..... ;
      in original form .....}

\pagerange{\pageref{firstpage}--\pageref{lastpage}}
\pubyear{2013}

\begin{document}

\maketitle

\label{firstpage}

\begin{abstract}

A model for massive stars is constructed by piecing together evolutionary algorithms for the protostellar structure, the environment, the inflow and the 
radiation feedback. We investigate specified accretion histories of constant, decelerating and accelerating forms
and consider both hot and cold accretion, identified with spherical free-fall and disk accretion, respectively.
 Diagnostic tools for the interpretation of  the phases of massive star formation and testing the evolutionary models are then developed.
Evolutionary tracks able to fit Herschel Space Telescope data require the generated stars to be three to four times less massive than in previous interpretations, thus being consistent with clump star formation efficiencies of 10 -- 15\%.  
However, for these cold Hershel clumps, the bolometric temperature is not a good diagnostic to differentiate between accretion models.
We also find that neither spherical nor disk accretion can explain the high radio luminosities of many protostars. Nevertheless, we discover a solution in which the extreme ultraviolet flux  needed to explain the radio emission is produced if the accretion flow is via free-fall on to hot spots  covering less than 10\% of the surface area. Moreover, the protostar must be compact, and so has formed through cold accretion.  We show that these conclusions are independent of the imposed accretion history. This suggests that massive stars form via gas accretion through disks which, in the phase before the star bloats, download their mass via magnetic flux tubes on to the protostar.
\end{abstract}


\section{Introduction}              

Massive stars are the principal source of heavy elements and ultraviolet radiation, and a 
major supplier of wind and supernova energy, within the Universe. Individually, they dominate cluster formation and,
collectively, they influence the evolution of galaxies. 
However, our knowledge of massive stars remains  limited  \citep{2007ARA&A..45..481Z}.
We know they are born in massive and dense clumps embedded within 
giant molecular clouds \citep{1991psfe.conf....3B}. We also think we understand fairly well how physical 
mechanisms conspire to describe the emergence of low- and intermediate-mass stars.  But the events which conspire 
to produce the high-mass counterparts are still controversial \citep{2007ARA&A..45..565M}. This is due to a combination 
of factors including their rarity,  large distances,  rapid evolution,  high extinction and their confusion with associated clusters. 

A formation model requires  integration of  evolutionary  models for the internal star, the environment
and the feedback.  Together, these constitute a    `protostellar system'  which consists of several distinct components including a hydrostatic core,
disk, envelope, wind, jets and bipolar outflow.  Recently, with Herschel, BLAST and Planck following Spitzer 
\citep[e.g.][]{2010A&A...518L..97E,2009ApJ...707.1824N,2011A&A...536A..22P}, we have acquired the quality and 
quantity of data to explore the early formation stages. 
Processes that  can now be addressed include  clump dispersal, cluster formation, mass inflow, stellar accretion, stellar flux and  mass outflow.

The aim here is to link the system components and underlying processes together through a model 
framework that predicts resulting observable correlations.
We thus develop a scheme which explores plausible paths from clump to exposed star.  It is set up to track how mass is passed between
the components, while accounting for direct ejection from the supplying clump and indirect ejection through jets.

The interaction of a gradually emerging star with a massive clump is described in terms of a time sequence
going from compact and hot molecular cores to extended H\,{\small II} regions \citep{2002ARA&A..40...27C} as expansion occurs.  This has emphasized an issue concerning the time scales with too many bound hypercompact or the later ultracompact  
 H\,{\small II} regions being observed. This may be resolved by taking less ideal approaches to accretion and dispersal from 
and to the clump \citep{2010ApJ...719..831P}, allowing for choking of the cavity through the infall of dense filaments and fragments of the clump.

The very early stages of protostar formation within cold clumps have now been identified in large numbers as Infrared
Dark Clouds (IRDCs) \citep{2006ApJ...653.1325S,2006ApJ...641..389R} .
These could be the star-less objects from which massive stars will form. However,
the earliest signs of accretion have also been signposted by directed outflows of cold molecular gas and, it turns 
out,  most of the IRDCs studied so far host weak 24$\mu$m emission sources and already drive molecular outflows,
both strong indicators for active star formation \citep{2010ApJ...715..310R}. In addition, the Herschel Space Telescope has helped reveal 
the full population of early evolutionary stages at the very onset of massive star formation \citep{2010A&A...518L..78B,2012A&A...547A..49R}.

It has been proposed that the large-scale evolution can be split into two phases: an accretion phase and a clean-up phase.
Initially, as the protostar gains in mass, its sphere of influence grows,  leading to
accelerated accretion \citep{2003ApJ...585..850M}. There is observational support for this as discussed by \citet{2011MNRAS.416..972D} which favours either turbulent core 
\citep{2003ApJ...585..850M} or  competitive accretion \citep{2001MNRAS.323..785B} models.  Subsequently, in a distinct second phase, termed the clean-up phase, the accretion has stopped abruptly and the remaining clump material is partly dispersed or integrated into a surrounding cluster \citep{2008A&A...481..345M}. 

In another scenario, the massive star does not fully  form early. Low mass star formation dominates until the clump has considerably 
evolved. These stars would remain difficult to detect. It has been remarked that the current Initial Mass Function implies two peaks 
of star formation with the majority of low mass stars forming first and high mass stars forming later \citep{2001A&A...373..190B}.
In competitive accretion, gas is funnelling down to the cloud centre where
stars, initially accreting gas with low relative velocity, already have large mass before accreting the late-arriving higher velocity gas \citep{2006MNRAS.370..488B}.

Stellar collisions and mergers could supplement  gas accretion  \citep{2002MNRAS.336..659B}. 
Massive stars could form via coalescence of intermediate mass stars within very dense systems. Although not considered here, along with other scenarios 
involving fragmentation, these  should lead to alternative predictions.

Do high-mass stars form in a scaled-up analogue of the low-mass formation scenario? \citet{2008A&A...481..345M} find a consistent 
interpretation in favour of an analogy. They analysed wide Spectral Energy Distributions (SEDs) to derive bolometric luminosity 
($L_{bol}$) and clump mass ($M_{clump}$). This leads to the diagnostic diagram of $L_{bol}$ versus $M_{clump}$ (we have replaced the term 
envelope with clump here, employing envelope to refer to the inner part of the clump which accretes directly on to the protostar). 
The high-mass objects were then shown to occupy a sequence of regions 
on this diagram in a similar manner to those evolving Classes which populate the low-mass regions. 
Herschel data have recently extended this to include protostars \citep[e.g.][]{2010A&A...518L..97E}. 

The SED parameters now available include fluxes derived from the Red MSX Source survey, Spitzer IRAC and MIPS, 
SOFIA FORCAST, Herschel PACS, Herschel SPIRE, and BLAST.
Further data are available across the radio, sub-millimetre and infrared. From these, we can derive  clump mass, temperature,  
luminosity, ultraviolet flux, outflow mass and outflow power. So we can now construct several diagrams to employ as diagnostic tools 
to estimate the evolutionary stages. 

The bolometric luminosity and temperature, $T_{bol}$, have been used in isolation to test accretion models for a version  of the model tuned to low-mass protostars \citep{2006MNRAS.368..435F}. A similar approach but just using the luminosity function was adopted by  
\citet{2011MNRAS.416..972D} for high-mass stars using MSX data and radio identifications to distinguish protostars from 
 later stages. In place of an SED derivation of luminosity, the 21\,$\mu$m flux was utilised with the assumption that the bolometric luminosity
 can be taken as a reliable proxy for the set of protostars being investigated. Both these approaches led to constraints on the time scale of 
 young stars and  the general form of the accretion process.

A potential test would use the radio luminosity, $L_r$, produced  by free-free emission after extreme ultraviolet excitation. The ratio of 
$L_r$/$L_{bol}$ should then provide a measure of the development while  $L_{bol}$/ $M_{clump}$ provides a distinct measure. Plotted together,
such a diagram provides a {\em distance independent} distribution of evolutionary phases. 

Outflow parameters have also been used to distinguish the phases of low-mass stars where accretion is known to decelerate 
rather than accelerate with time \citep{2006A&A...449.1077C}. This leads to a decrease 
in the force of the outflow as the source ages \citep{2010MNRAS.408.1516C}.
 \citet{2002A&A...383..892B} showed that bipolar outflows are
indeed  ubiquitous phenomena in the formation process of massive stars, suggesting similar flow-formation scenarios for all masses,
consistent again with scaled-up, but otherwise similar, physical processes - mainly accretion - to their low-mass counterparts. Going further,
it was shown that the measured molecular hydrogen outflow luminosity is tightly related to the source bolometric luminosity for 
low mass stars \citep{2006A&A...449.1077C},  and this relationship extends to massive objects \citep{2008A&A...485..137C}.  
It is clear that this only applies to the youngest protostars in which the bolometric luminosity is dominated by the 
release of energy through accretion. Those sources associated with jets are very young (well before the Main Sequence  turn-on), while 
those without detectable jets  possess ultracompact  H\,{\small II} regions \citep{2005ccsf.conf..105B,2011BSRSL..80..235R}.

The objectives of this first paper is to set up the model framework and consider the effects of radiation feedback. We impose  simple accretion
rates that slowly vary.  Subsequent works will tackle  accretion outbursts, the outflow properties and the strength of feedback in  self-regulation.

The accelerated-accretion model generates a particular $L_{bol}$ versus $M_{clump}$ relation  \citep{2008A&A...481..345M}. We begin here by exploring how general this is.  A constant accretion rate is a common working assumption while there is evidence for both a declining rate as well as a sporadic/episodic rate. These models involve less dramatic 
cut-offs in the accretion and should generate models with different statistical properties. 

A major issue to address later is the existence of two distinct phases.
If accelerated accretion is followed by a clean-up phase, we expect the outflow feedback phase to precede the radiative feedback phase.
It may prove difficult for accretion phenomena to still dominate once the rapid rise 
in ultraviolet flux has started at late times. Quite remarkably, nature has no problem:  all of the sources with infall signatures onto
Ultracompact H\,{\small II} regions have corresponding  outflow signatures as well \citep{2011arXiv1112.0928K}. This observation
suggests that accretion may continue, consistent with the gravo-turbulent model \citep{2004A&A...419..405S}. However,
both accretion and collimated outflows are probably weak when the star has advanced to its Ultracompact H\,{\small II} 
stage \citep{2010MNRAS.404..661V}. An ultimate aim of this study will therefore be to determine the conditions under which  the accretion-outflow phase can significantly
overlap with the UV phase, associated with the final contraction of a massive star.

\section{Method}              
\label{method}

\subsection{ Model construction: mass movement}
\label{model}

The  model is constructed upon (1) the extension of the Unification Scheme for low-mass stars, 
(2) the strategy and model invoked for high-mass stars, (3) the detailed evolutionary tracks of an accreting massive protostar, 
(4) the results for a range of potential accretion rates and (5) predicted outcomes for radiative and outflow feedback. All algorithms
and graphics are written and processed in IDL.

The first task is the construction of a model for the environment in which a given clump mass, $M_{clump}$, is redistributed
in time according to a prescribed formula, constrained by mass conservation. The clump directly supplies three entities: an 
inner envelope. $M_{env}$, a surrounding cluster  $M_{stars}$ and dispersal into the ambient cloud $M_{gas}$ (with 
some additional help from the jet-driven outflow). The inner envelope is here assumed to supply the accretion disk at a rate  
$\dot  M_{acc}(t)$ which, feeds instantaneously both the star, $M_*$, and the jets, $M_{jets}$. The jet material accumulates 
in an inner outflow, $M_{out}$, which also pushes out a fraction of the clump.  Hence:
\begin{equation}
   \dot  M_{clump}(t) +  \dot  M_{env}(t)  + \dot  M_{stars}(t) + \dot  M_{gas}(t)  = 0,
\end{equation}
\begin{equation}
   - \dot  M_{env}(t) =   \dot  M_{acc}(t) = \dot  M_{*}(t) + \dot  M_{jets}(t) .
\end{equation}

We will consider two important  free parameters. The most critical is the fraction, $\xi$, of the initial clump mass which ends 
up as part of the star. Even for low mass stars, it is well known that there must be a much larger obscuring mass than necessary 
to form the star \citep{1998ApJ...492..703M}. This mass  prolongs the embedded phase and extends the late Class 0 and early Class 1 stages. The best estimate for the excess mass was found to be a factor of two in the low-mass version of the present scheme on using accretion rates derived from gravo-turbulent models \citep{2006MNRAS.368..435F}.  

Also for high-mass protostars, the surrounding bound clumps are estimated to exceed the star's  mass  by a factor which can exceed 30 \citep{2008A&A...481..345M}. We thus anticipate quite low values for  $\xi$.  Hence, in this work, we assume that the clump mass is sufficient to form the protostar and the associated stellar cluster (see Sub-section~\ref{clumps}) in addition to gas expelled directly from the clump. We assume both mass loss rates to be constant with the same time scale. 

The second parameter is the fractional efficiency $\epsilon$ of  mass diversion from inflow to outflow,  from the disk to the jets. The extended magneto-centrifugal 
model is expected to be quite efficient and the X-wind model is expected to reach the thirty per cent  level  \citep{1988ApJ...328L..19S,1994ApJ...429..781S}. In doing so, 
such magneto-centrifugal  mechanisms can carry away the total angular momentum and kinetic energy of the accreting disk material. 

Previously, we found that a constant efficiency  $\epsilon$ was inconsistent with the  observations of low-mass stars \citep{2000IrAJ...27...25S}.  Hence we took
\begin{equation}
     \epsilon(t) = \eta \left[\frac{\dot M_{acc}(t)}{\dot M_o}\right]^\zeta,
\label{eqn-hm}
\end{equation}
where $\dot M_o$ is the peak accretion rate and $\zeta$ is a constant. 

The variable jet efficiency was introduced in order to account for evolving outflow properties of low-mass protostars. The Class 0 outflows appear to have 
a mechanical luminosity of order of the bolometric luminosity of the  protostellar core. On the other hand, Class 1 outflows have mechanical 
luminosities and momentum flow rates up to a factor of 10 lower \citep{1996A&A...311..858B,2006A&A...449.1077C}. 
With the same mass outflow efficiency, $\zeta = 0$, this would then require that the  {\em jet speed} is higher in Class 0 
outflows. This is, however, contrary  to the observations which  associate lower velocities  ($\sim 100$\,km\,s$^{-1}$) to 
the Class 0 outflows.  Here, we shall assume a constant  outflow mass fraction,   taking the case $\epsilon = 0.3$ throughout this work.

In general, the mass left over which accretes on to the core to form the star is
\begin{equation}
M_*(t) = \int_0^t  (1\,-\,\epsilon(t))\,\dot M_{acc}(t)  dt.
\end{equation}

\subsection{ Model construction: accretion rates}

\begin{figure}
\includegraphics[width=8.7cm]{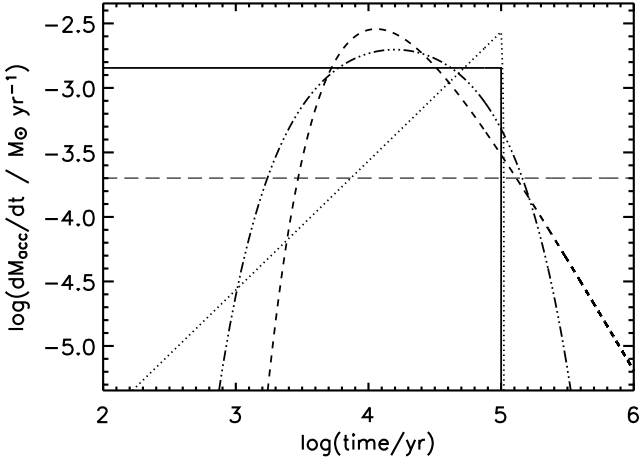}
\caption{Accretion rates as a function of time (log-log plot). The rate of loss of mass from the envelope is displayed for the five models discussed in the text. In each case a star of mass 100~M$_\odot$  is formed on assuming 30\% of the envelope mass is ejected in jets. The models are (i) constant fast accretion for 10$^5$~yr
(solid line), (ii) constant slow  accretion for $\times$~10$^6$~yr (long-dashed),  (iii)  power law with maximum $\dot M_{acc}$ = 4.01 ~$\times$~10$^{-3}$~M$_\odot$~yr$^{-1}$ (dashed), (iv) exponential with maximum $\dot M_{acc}$ = 2.77 ~$\times$~10$^{-3}$~M$_\odot$~yr$^{-1}$ (dot-dashed), and (v) accelerated accretion for  10$^5$~yr
(dotted  line). \label{rateversustime}}
\end{figure}

The mass accretion rate $\dot M_{acc}(t)$ is the main prescribed parameter. We choose the four forms as shown in Fig.~\ref{rateversustime} 
and set out below. Note that the power-law and exponential models both include a significant phase of accelerated accretion prior to 
the prolonged decline.

\subsubsection{Constant rate models}

Constant accretion models were favoured following the work of \citet{1977ApJ...214..488S} on singular isothermal spheres. It is clear,
however, that the rate must eventually fall (by the pre-main sequence stage for low-mass stars) as the reservoir becomes
exhausted. Here we assume a constant rate until a cut-off time, $t_o$, at which the accretion is abruptly halted.

More generally, the collapse of isothermal isotropic cores  supported by thermal pressure yields infall rates of 
the form $\dot M_{acc} = f(t)\left[c_s^3/G\right]$ where $c_s$ is the sound speed \citep{1993ApJ...416..303F}. 
The function $f(t)$  can be  anything from  gradually decreasing over a time $\sim\,200\tau$ 
to a sharply peaked function at time $\sim\,0.1\tau$ where the time and mass flow scales are
\begin{equation}
    \tau = 1/{\surd}(4{\pi}G\rho_c) = 1.3\,10^5 
           \left[\frac{7.6\,10^{-20}\, {\rm g\,cm}^{-3}}
           {\rho_c}\right]^{1/2} {\rm yr}
\end{equation}
and
\begin{equation}
     \frac{c_s^3}{G} = 1.6\,10^{-6} \left[\frac{c_s}
     {0.19\,{\rm km\,s}^{-1}}\right]^3 {\rm M}_{\odot}\,{\rm yr}^{-1},
\end{equation}
where $\rho_c$ is the initial central density.  Inclusion of envelope spin, turbulence, magnetic field and fragmentation  
would introduce further initial parameters (see review by \citet{2003RPPh...66.1651L}).

\subsubsection{Exponential models}

 One type of model assumes the envelope loses mass in proportional to its mass,
$\dot M_{acc}\,=\,-\dot M_{env}\,\propto\,M_{env}$ \citep{1996A&A...311..858B,1998ApJ...492..703M}. This
leads to an exponentially decreasing envelope mass 
\begin{equation}
   M_{env}(t)   =   M_o\,e^{-t/\tau_f},
\end{equation}
and an accretion rate
\begin{equation} 
   \dot  M_{acc}(t) =  \frac{M_o}{t_f}\,e^{-t/\tau_f},
\end{equation}
which thus begins with an established high accretion rate.
A modified exponential model allows for a rapid rise and fall by modelling
\begin{equation}
   \dot  M_{acc}(t) = 
   \dot  M_o\,e^{2\left(\tau_r/\tau_f\right)^{1/2}}\,
                 e^{-\tau_r/t}\,e^{-t/\tau_f},
\label{eqnexp}
\end{equation}
and we can put $\tau_r = 0$ if desired. The maximum accretion rate $\dot M_o$ 
occurs at time $t_m = {\surd}(\tau_r\tau_f)$.

\subsubsection{Power law models}

The favoured model for low-mass evolution involves a sharp exponential rise  followed by a prolonged
 power law decrease in time \citep{1999ASPC..188..117S,2000IrAJ...27...25S}. The power-law has substantial observational
support \citep{2000prpl.conf..377C}. The early peak may reach 
 $\dot M_{acc} = 10^{-4}M_{\odot}\,{\rm yr}^{-1}$ for 10$^4$ years, and eventually fall to 
 $\dot M_{acc} = 10^{-7}M_{\odot}\,{\rm yr}^{-1}$ for 10$^6$\,years, corresponding to Class 0 and Class 2 or 
 Classical T~Tauri stars, respectively. 

We choose the accretion rate to take the form
\begin{equation}
   \dot  M_{acc}(t) = 
   \dot  M_o \left(\frac{e}{\alpha}\right)^{\alpha} \left(\frac{t}{t_o}\right)^{-\alpha} e^{-t_o/t}.
\label{eqnpower}
\end{equation}
Note that  we can choose $\alpha$ to simulate various models: an asymptotic constant-accretion model corresponds to $\alpha \sim 0$ and $t_o$ small,
gradual-accretion corresponds to $\alpha \sim 0.5$ and abrupt accretion to $\alpha \sim 2-3$. In this work, we take  $\alpha = 1.75$ as a default value 
The envelope evolution can be written analytically in terms of an incomplete
Gamma function on integrating Eqn.\,\ref{eqnpower}:
\begin{equation}
   M_{env}(t) = 
   \dot  M_ot_o (e/{\alpha})^{\alpha}\left[1\,-\,{\Gamma}
                     (\alpha-1,t_o/t)\right].
\end{equation}

\subsubsection{ Accelerated accretion}

A power-law form for the accretion rate, in which the star's growth accelerates, takes a simple form:  
\begin{equation}
   \dot  M_{acc}(t) =   \dot  M_o \left(\frac{t}{t_o}\right)^{n},
\label{eqnaccel}
\end{equation}
for $t  <  t_o$ with the final, maximum rate $\dot M_o$. The accretion in this case is assumed to be
 $\dot M_{acc} = 0$ for  $t  >  t_o$. This yields a final mass for the star of
 \begin{equation}
   M_f =   \frac{\dot M_o t_o (1-\epsilon)}{1+n},
\label{eqnmasspower}
\end{equation} 
assuming $\epsilon$ is constant.

The sudden drop in the accretion rate at $t  =  t_o$ appears as 
a sharp spike in $L-M$ plots since the contribution to the bolometric luminosity from the accretion luminosity
disappears.  This can be seen in the tracks calculated by \citet{2008A&A...481..345M}. Here,
we include a brief period of linear decline down to a minimum accretion rate of $10^{-4}$~$\dot M_o$: 
\begin{equation}
    \dot  M_{acc}(t) =  20 \times  \dot M_o ( 1.05 - t/t_o) \label{eqnmassadjust}
\end{equation} 
for $t_o < t < 1.05t_o$. 

The turbulent core model is reproduced with $n = 1$ which leads to
$\dot M_{*} \propto   M_{*}^{1/2}$. This model has been shown to lead to some predictions which are 
consistent with different sets of data \citep{2008A&A...481..345M,2011MNRAS.416..972D}.

Note that observed correlations between accretion rates and mass such as
$\dot M_{acc} \propto M_{*}^{1.8 \pm 0.2}$ \citep{2006A&A...452..245N} apply to a
relatively mature stage of  young stars.  In contrast, $\dot M_{acc} \propto M_{*}^{1}$ was
 uncovered \citep{2011MNRAS.415..103B,2012MNRAS.421...78S}. However, it is no surprise that
the results differ according to the precise sample selection criteria 

\subsection{ Growth of the star}

The structure of a protostar while it accretes at a constant rate has been calculated by \citet{2009ApJ...691..823H} in the
`Hot Accretion' scenario corresponding to spherical free-fall. If, instead of free-fall on to the surface, the gas settles via an 
accretion disk, the  `Cold Accretion' structure is appropriate \citep{2010ApJ...721..478H}.
In the Hot Accretion case, the stellar radius swells up to over 100~$R_\odot$ for $\dot M_{*} > 10^{-3}$~M$_\odot$~yr$^{-1}$.
The accretion may continue until after the arrival on the Main Sequence, arriving at higher masses for higher accretion rates.

\begin{figure}
\includegraphics[width=8.7cm]{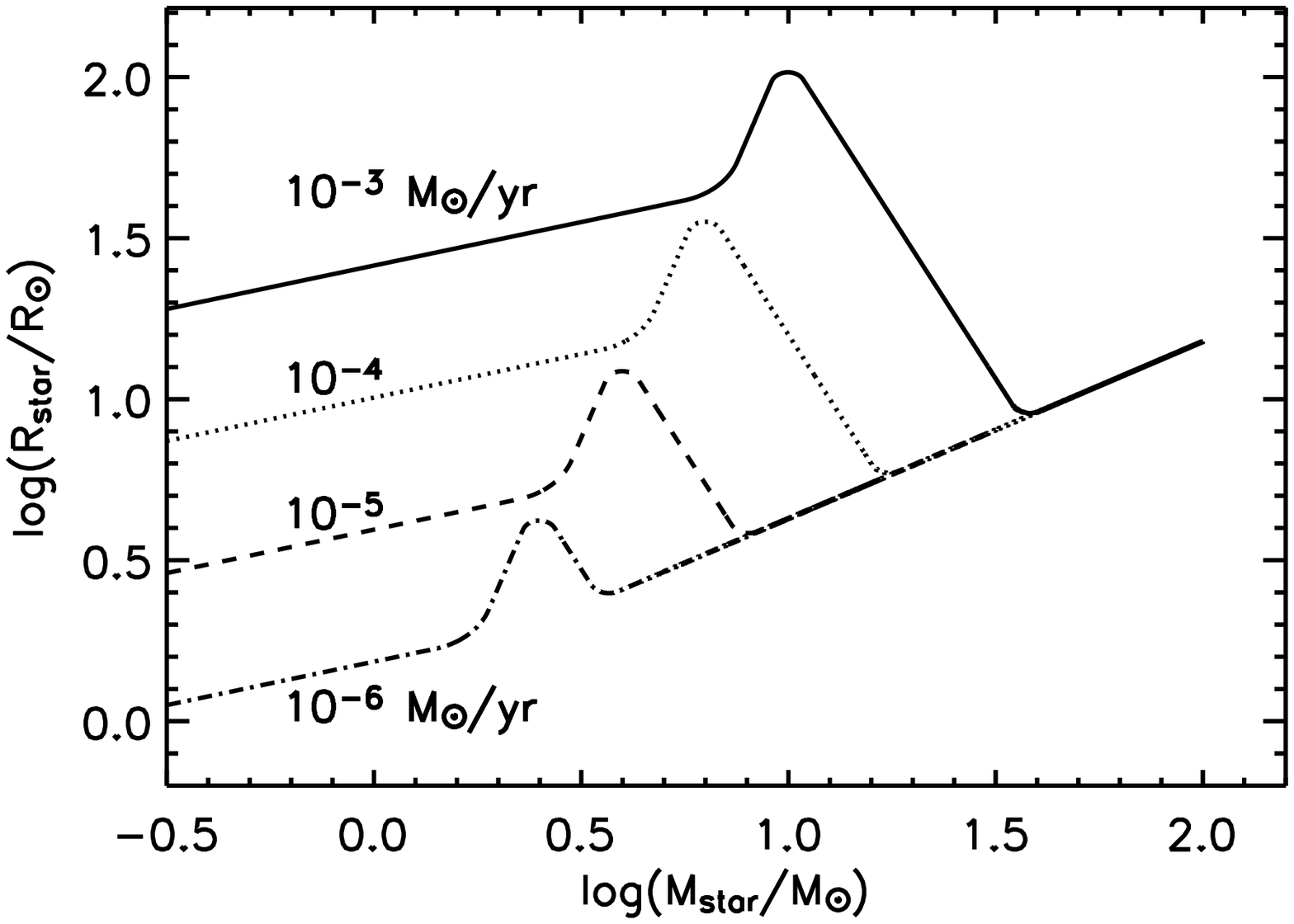}
\includegraphics[width=8.7cm]{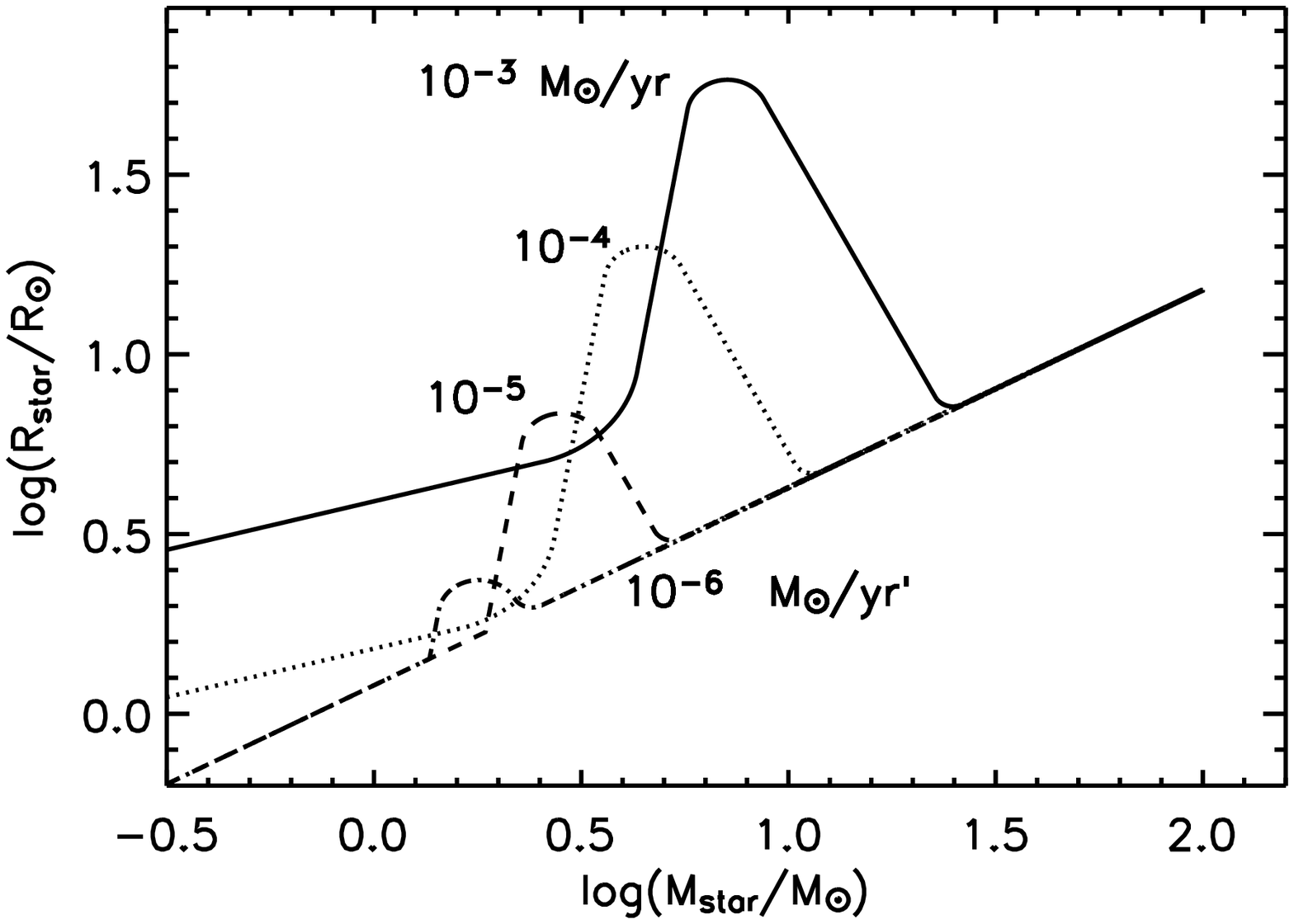}
\caption{Stellar radius evolution for Hot Accretion (upper panel) and Cold Accretion (lower panel).. The mass is accreted for four constant accretion rates, $\dot M_*$, as indicated. These were calculated from analytical approximations to the data presented by \citet{2009ApJ...691..823H}   for the case of  hot accretion and
\citet{2010ApJ...721..478H}   for the case of cold accretion.  \label{radius}}
\end{figure}

\begin{figure}
\includegraphics[width=8.7cm]{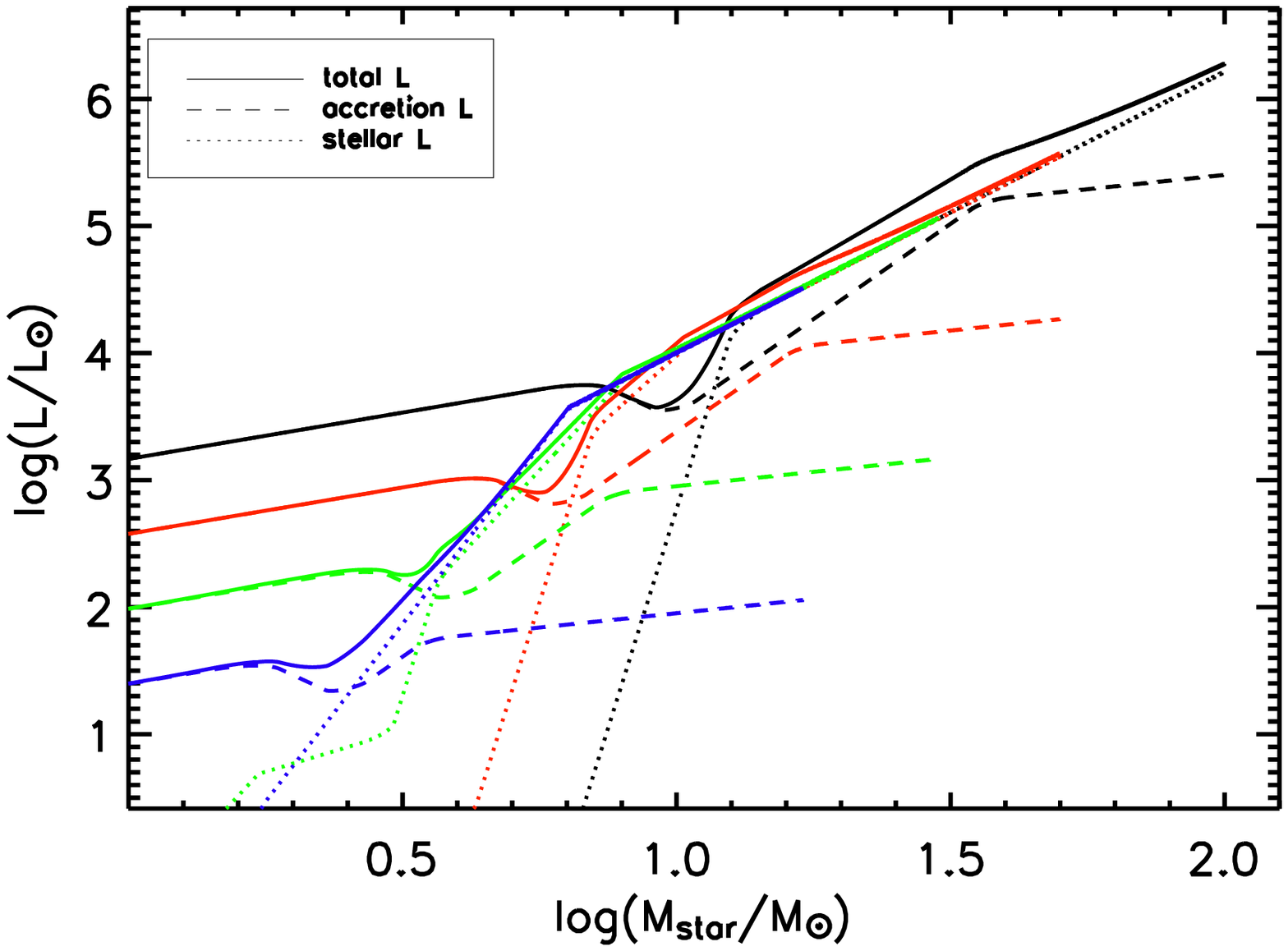}
\includegraphics[width=8.7cm]{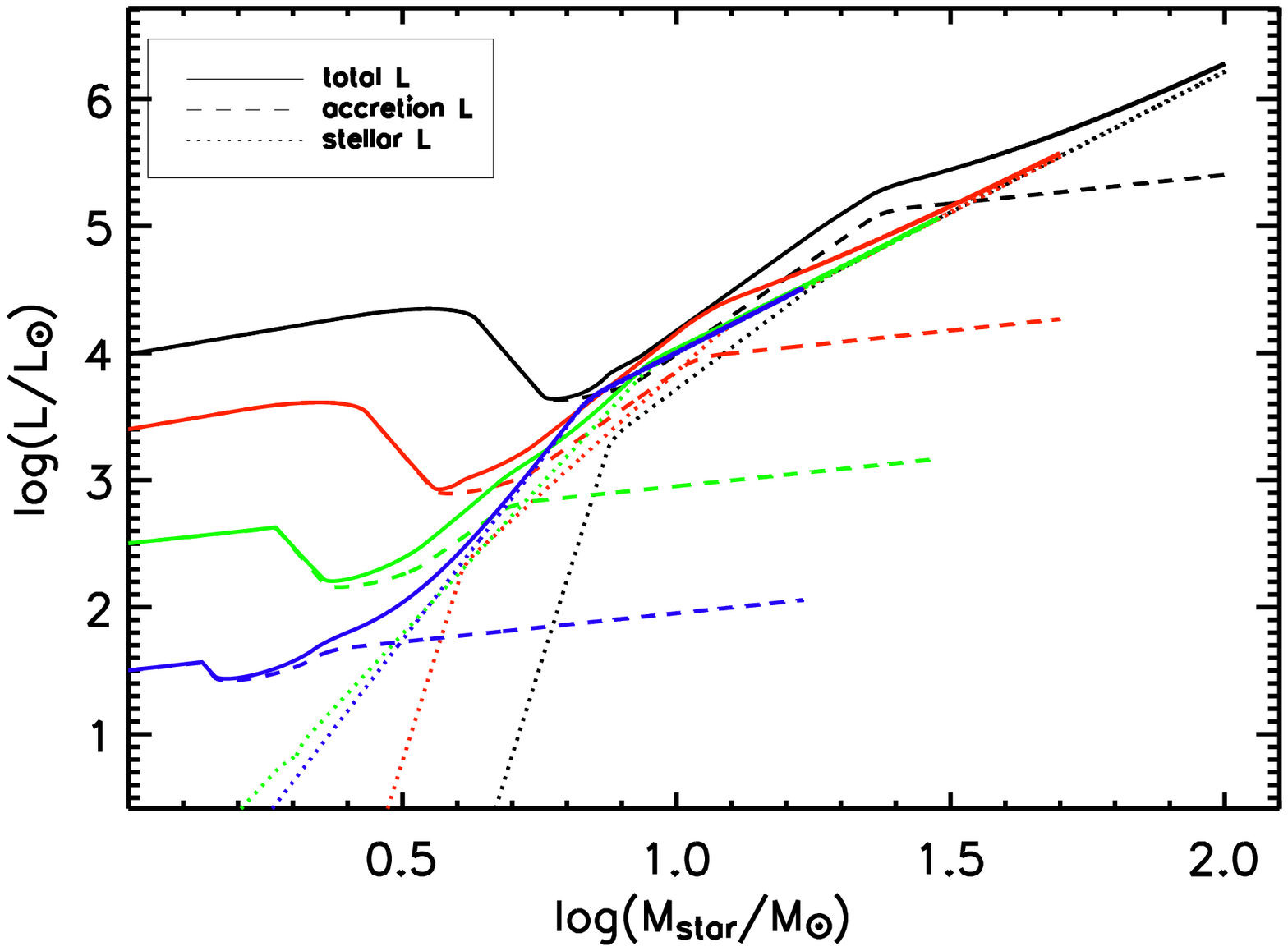}
\caption{Luminosity evolution for Hot Accretion (upper panel) and Cold Accretion (lower panel). The total (solid), accretion (dotted) and stellar (dashed) luminosities are displayed for four constant accretion rates, $\dot M_*$, creating stars of final mass $M_f$ with $\dot M_* = 10^{-3} $M$_\odot $yr$^{-1}$ and $M_f = 100 $M$_\odot $ (black lines), $\dot M_* = 10^{-4} $M$_\odot $yr$^{-1}$  and $M_f = 50 $M$_\odot $ (red lines), $\dot M_* = 10^{-5} $M$_\odot $yr$^{-1}$  and $M_f = 30 $M$_\odot $ (green lines),
and $\dot M_* = 10^{-6} $M$_\odot $yr$^{-1}$  and $M_f = 17 $M$_\odot $ (blue lines),
calculated from analytical approximations to the data presented by \citet{2009ApJ...691..823H}   (hot accretion)
and \citet{2010ApJ...721..478H}    (cold accretion).  \label{luminosity}}
\end{figure}

For this work, we have fitted analytical functions to the template figures provided in the above two studies for the radius, $R_*$, and luminosity, $L_*$.
The four functions correspond to the four main stages with 
smooth interpolation between these stages. The stages are (1) the adiabatic accretion, (2) the swelling (or bloating), (3) the Kelvin-Helmholtz contraction and, finally, (4) main-sequence accretion. The resulting functions are shown in Figs. \ref{radius} and \ref{luminosity}.  Figure \ref{luminosity} displays both the accretion and stellar luminosities, and demonstrates that the accretion luminosity dominates until the radial swelling stage which is apparent as a dip in the bolometric luminosity.

The stellar structure depends on the stellar mass,  the initial interior state and the accretion rate history. For the case of
 a constant accretion rate, the above published  works provide accurate  templates  for fiducial cases. However, to employ these figures for time-varying accretion,
 \citet{2011MNRAS.416..972D} took the current  accretion rate to look up the radius and luminosity from \citet{2009ApJ...691..823H}.  This method
may be a reasonable approximation when the accretion rate continues to increase such that most of the mass has been accumulated 
within a factor of two of the present accretion rate. 
More accurately for the adiabatic phases, we here calculate how the star has accumulated the mass and entropy over its entire evolution. Thus a mass-averaged accretion rate,  proportional to the accumulated entropy, is employed.    
 However, this method would still lead   to very large errors when the accretion rate varies by large amounts, decreases considerably or varies rapidly in the post-adiabatic stages.
   

When the above method yields a  Kelvin-Helmholtz time that is comparable or shorter than the  accretion time scale, the adiabatic  approximation is invalid. 
A compromise solution would employ both the current and the mass-avaeraged accretion rates so as to deliver the correct stellar parameters for the two limiting cases of adiabatic accretion and rapid loss of entropy. This is parameterised  with
the use of the factor  $f = t_{acc}/t_{KH}$ where, as usually defined, $t_{acc} = M/\dot M$ and $t_{KH} = G M^2/(R L_{int})$, as determined by the mass-averaged accretion rate. We then take this factor to determine the stellar radius and luminosity relative to the mass averaged values, $R_o$, $L_o$, by the additive factors
 $(1 - exp(-f)) \times (R(t) - R_o)$ where $R(t)$ is the stellar radius calculated from the current rate.   While this compromise is more accurate, this would still lead to spurious
 properties if accretion variations are very rapid.
 
In this paper, therefore, we will use a second method, where the structure at any time is determined by the mass-averaged accretion rate over the history of the star limited to the past Kelvin-Helmholtz time. We implement an iterative process to determine the radius, luminosity and mass-average accretion rate since these are themselves functions of $t_{KH}$.  Upon testing, we find that the maximum difference between the two methods is typically a few per cent and occurs in the Kelvin-Helmholtz phase. In comparison, simply assuming the current accretion rate and reading the stellar structure from the constant-rate simulations, generated errors of order of 10\% in radius throughout the first three evolutionary stages for the slowly varying rates displayed in Fig.~\ref{rateversustime}.  Although not ideal, it  could still be implemented for most present purposes.


\subsection{Radiation feedback modelling}
\label{radiation}

The bolometric luminosity, $L_{bol}$, and the jets' mechanical power, $P_{jets}$, are equated to the 
gravitational energy released through accretion and the stellar luminosity:
\begin{equation}
   L_{bol} +  P_{jets} =   L_{acc} + L_* +  P_{jets} = \frac{M \dot  M_{acc}}{R_*}  + L_* .
\end{equation} 

The jets' power is related to the jet mass outflow and accretion rates by 
\begin{equation}
        P_{jets} = \frac{1}{2}\dot  M_{jets} v_{jet}^2 = \frac{1}{2}{\epsilon}\dot  M_{acc} v_{jet}^2.
 \end{equation}  
 We take
 \begin{equation} 
        v_{jet}^2  = \chi^2 \frac{G M_*}{R_*},
 \end{equation}  
 with $\chi = 1$ as the default constant value, corresponding to a jet speed proportional to the escape speed from the stellar surface.

The accretion luminosity is thus
\begin{equation}
     L_{acc} =  (1- \frac{1}{2}\chi^2{\epsilon})   \frac{M \dot  M_{acc}}{R_*}  .
\end{equation}  
It should be noted that this remains an approximation since various factors, such as stellar rotation,
are not accounted for.

The radiation feedback into the environment consists of an ionizing effect and a heating effect on the surrounding envelope.
 The ultraviolet Lyman flux from the protostar and the accreting material ionizes the environment, generating an H\,{\small II} region. 
  This region can be observed in the radio continuum through free-free emission. 
  
  The Lyman flux was tabulated in the work of \citet{2011MNRAS.416..972D} for hot main sequence stars. This provides an excellent up-to-date blueprint
  for stars evolving on to the Main Sequence although expected uncertainties remain very high. 
  Since they considered that significant Lyman  flux in their sample would only occur when the star was as good as on the Main Sequence, they were able 
  to directly use their tabulated values given the stellar mass. In the present study, we have calculated approximations to the Lyman flux based on the work of
  \citet{1973AJ.....78..929P}  and \citet{2011MNRAS.416..972D} to relate surface temperature to the Lyman flux, and then employed the stellar temperature and radius  to determine the flux of Lyman photons. We find that the two works are in close agreement and power-law approximations appropriately
represent them   In detail, the fit used is  $L_{Ly} = 4 \pi R_*^2 F_{Ly}$ with
    \begin{equation}
       log( F_{Ly}) =  23.14 + 15.8(log(T_*) - 4.5 )     ~~~~~~log(T_*)  < 4.55
    \end{equation}     
 \begin{equation}
       log( F_{Ly}) =  24.7 + 5.1(log(T_*) - 4.7 ),  ~~~~~~log(T_*)  > 4.55. 
   \end{equation}        
      
To convert a predicted  ultraviolet Lyman flux into a high-frequency radio flux  density is straightforward. Provided the observed region is 
ionisation bounded and the frequency is sufficiently high to ensure that self-absorption is negligible, then the two are approximately proportional
 \citep{1968ApJ...154..391R,2013A&A...550A..21S}:
  \begin{equation}
      \frac{N_{Ly}}{ counts~~s^{-1}}  = 43.6  \left(\frac{S_{5GHz}}{mJy}\right) \left( \frac{D}{ kpc }\right)^2  ,
  \end{equation}     
which yields
 \begin{equation}
      \frac{L_{Ly}}{ L_\odot} = -0.63 \left(\frac{S_{5GHz}}{mJy}\right) \left( \frac{D}{ kpc }\right)^2  .
  \end{equation}   
  
\subsection{Envelope \& bolometric temperature}
\label{envelope}

The observed spectral energy distributions of protostars  are often complex with   multiple peaks. In this work, we restrict the analysis to 
predicting the bolometric temperature of the optically-thick core of the clump given a spherically-symmetric model. 
This provides an indication of how the peak wavelength will change with age. However, this should be only considered indicative since, even in this 
homogeneous spherically symmetric approximation, we  require several other major assertions concerning the gas and dust distribution.
  
 We consider two methods corresponding to different  clump detection and measurement scenarios. In Method 1, we first set up a large clump  of mass 
$M_{clump}$ with a mass related to the most massive star (see Equation~\ref{clumpmass} below) with a radial power-law density distribution.
This clump model was explored by \citet{2006MNRAS.368..435F}, and the sensitivity to parameter ranges was tested. Following those results, we take 
a small fixed inner envelope radius, $R_{in}$, of 30\,AU, in order  to sustain a high accretion rate.  We calculate an optical depth of the clump $\tau_c$ 
assuming a spherical density structure with $\rho \propto R^{-\beta}$.  The outer clump radius, $R_{out}$, is taken to be located where the clump merges into
the ambient cloud, i.e. where the temperature has fallen to that of the ambient molecular cloud, taken here as 12\,K. The inner clump density is then   
\begin{equation}
     \rho_{in} = \frac{ (3 - \beta )M_{clump}}{4{\pi}R_{in}^3\zeta }
\label{eqnrhoe}
\end{equation}
where $\zeta = (R_{out}/R_{in})^{(3-\beta)} - 1$, and the clump  optical depth is
\begin{equation}
    \tau_c =  \kappa\,\rho_{in}R_{in} \frac{ 1-\zeta^{1-\beta} } {\beta - 1}.
\label{eqntaue}
\end{equation}
We follow Myers et al (1998) and take the emissivity at 12$\mu$m as $\kappa = 4\,{\rm cm}^2\,{\rm g}^{-1}$. 

The inner temperature of the clump is
\begin{equation}
     T_{\rm in}(t) = \left[\frac{L_{bol}(t)}{4 \pi \cdot \sigma \cdot    R_{in}^2}\right]^{1/4}. 
\end{equation}
This yields an optical depth through the envelope proportional to the emissivity: 
\begin{equation}
   \tau_{c}(t) = \kappa \cdot \rho_{in}(t) \cdot R_{in} \cdot \frac{1 -   \zeta(t)^{1 - \beta}}{\beta - 1}. 
\end{equation} 
Following \citet{1998ApJ...492..703M}, we take$A = 1.59~10^{-13}$~cm$^2$\,g$^{-1}$\,Hz$^{-1}$, $h$
as the Planck constant, $k$ the Boltzmann constant, and calculate the bolometric temperature from
\begin{equation}
     T_{bol}(t) = \frac{\Gamma(9/2)\cdot \zeta(9/2)}{\Gamma(5)\cdot \zeta(5)}
     \cdot \left[\frac{h \cdot \kappa \cdot T_{in}(t)}{k \cdot A \cdot
     \tau_{e}(t)}\right]^{1/2}. 
\label{tbolometric}
\end{equation}

\begin{figure}
\includegraphics[width=8.7cm]{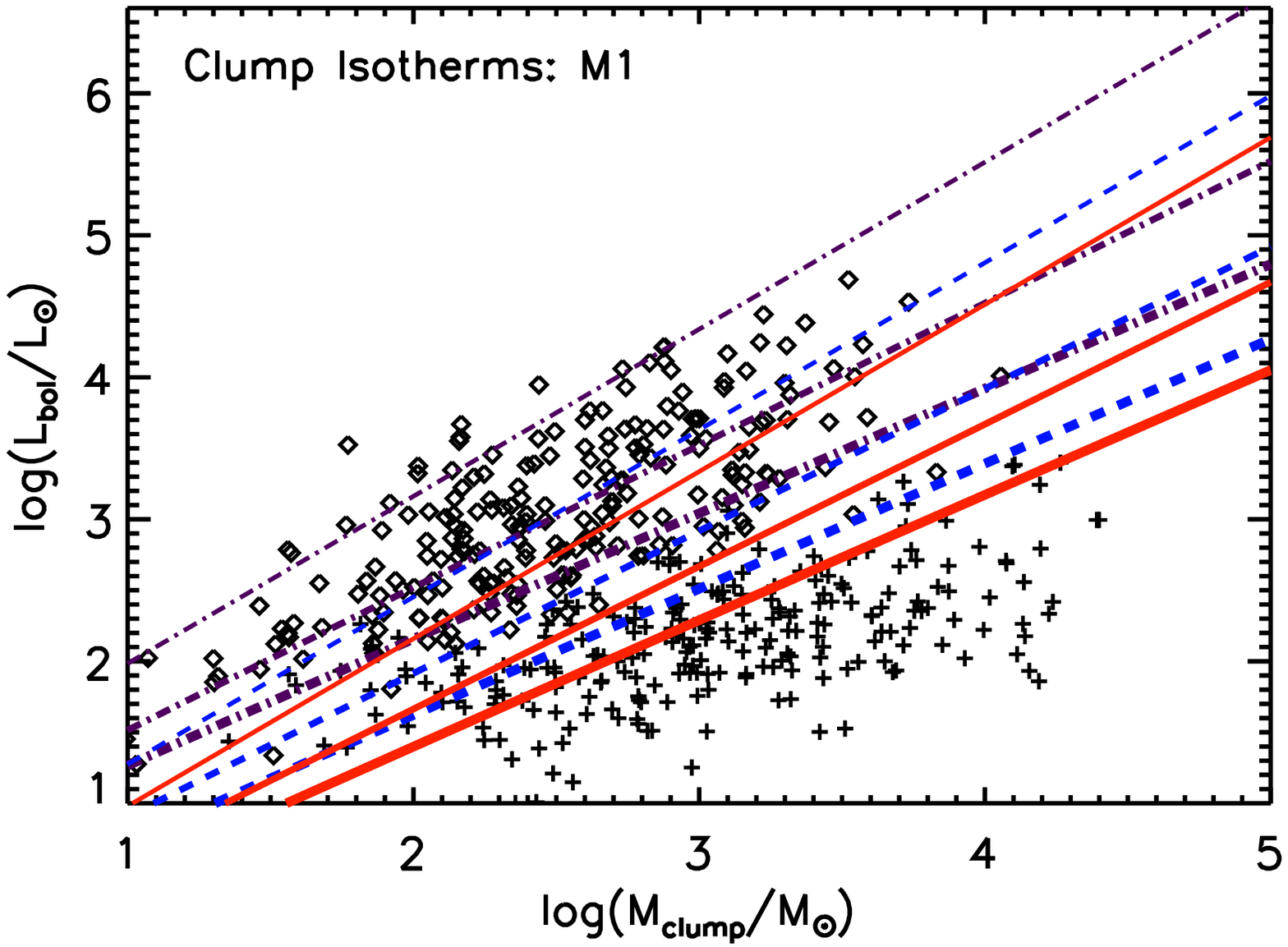}
\includegraphics[width=8.7cm]{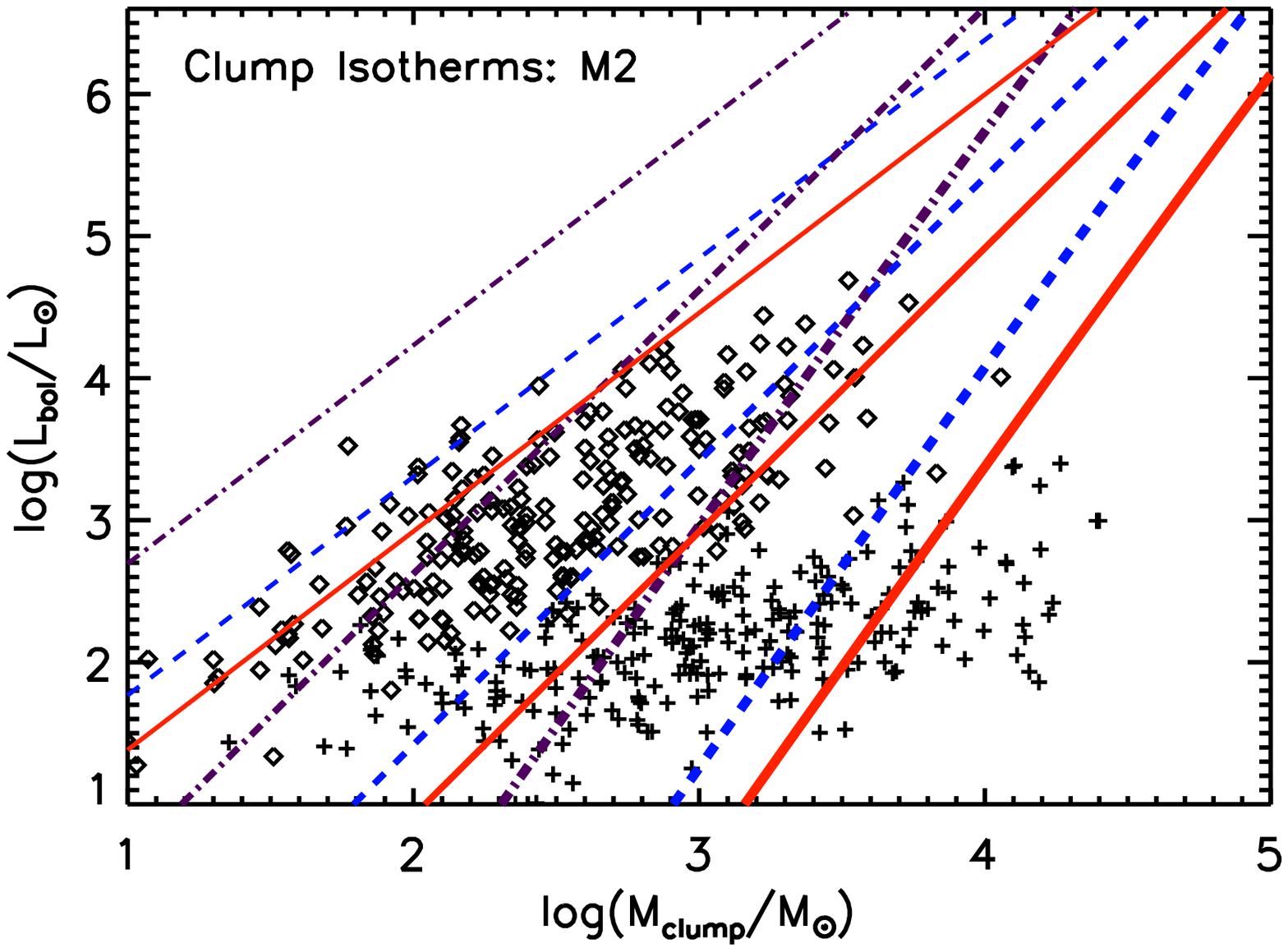}
\caption{Isotherms for the bolometric temperature of clumps plotted on the L$_{bol}$--M$_{clump}$  diagram. The upper panel employs Method 1 with a fixed inner radius of 30\,AU
and an outer radius of temperature 12\,K while Method 2 is displayed in the lower panel where the inner radius is determined by dust sublimation and the outer radius is a constant of 
50,000\,AU (i.e. full size of 0.5\.pc). The isotherms are for temperatures of 15\,K (full red), 30\,K (dashed blue) and 45\,K (dot-dash black) with $\beta$ = 1.2
(lower thick), 1.5  and 1.8 (upper thin). 
The data are from the Herschel $l = 30^\circ$ field  as analysed  by \citet{2013A&A...549A.130V} with protostellar clumps (diamonds) and pre-stellar clumps (crosses,
taken as clumps with no detected stars but which are gravitationally bound if the temperature is used to derive the relevant internal measure of  pressure).\label{isotherms}  }
\end{figure}

The above Method 1 was devised for the low-mass protostellar case in which the clump is described as a core, and its mass  may correspond to that of 
the enhanced density and temperature which distinguishes it from the embedding molecular cloud. This case leads to linear isotherms on the diagnostic  
L$_{bol}$--M$_{clump}$ logaritmic diagrams with the bolometric temperature simply proportional to M$_{clump}$ / L$_{bol}$ for the case $\beta = 1.5$. 
As shown in the top panel of  Fig.~\ref{isotherms}, the predicted isotherms are quite close together thus requiring a wide range of bolometric temperatures to cover the data 
for the displayed Herschel clumps.

In the high-mass case, we associate the clump with a size that is only mildly dependent on the clump mass. The observed median half-size is $ 5 \times 10^4$~AU 
\citep{2008A&A...481..345M,2013A&A...549A.130V}. Here, we will assume this value as a constant outer radius with larger clumps  having the extra mass `squeezed in'. as concluded by 
\citet{2013A&A...553A.115B}.  To complete the model for Method 2, the inner radius, R$_{in}$, of the clump is taken to be the sublimation radius  i.e. T$_{in} = 1,400$\,K. 
With these two boundary conditions replacing those of Method 1 (the distribution otherwise the same as described above), this Method 2 yields
well-separated isotherms as shown in the lower panel of Fig.~\ref{isotherms}. We will display isotherms derived from Method 2 in the following.

\subsection{Disk accretion}

At the inner radius, the envelope feeds  a circumstellar disk. We assume here, and will test in a following study, the working assumption that the disk is `viscous'
and steady, and that the gas spends relatively little time in the disk and so reaches an inner accretion disk at the same rate as with which it is supplied by the envelope. 
The inner accretion may be non-steady, the gas either being expelled in the jets or accumulated onto the surface of the protostar, later to become the star itself. The disk mass is thus proportional to the accretion rate and the accretion time scale.

Standard turbulent  viscosity is efficient at separating the flux of angular momentum  from the mass  out to radii of about 100\,AU for the initial rapid 
accretion rates from the envelope. Hence massive outer disks could build up until the viscous mechanisms associated with 
self gravity are effective. This could lead to the formation of secondary objects (stars, brown dwarfs), and so cut off both the star and jet supply
line. Perhaps more likely is that high accretion rates lead to simultaneous binary formation and powerful molecular jets.

We assume here that the inner disk processes the material fast and so remains steady. The outer part of the disk will lag behind. The outer radius of the
steady state disk can be found by requiring the disk accretion time scale $t_{\nu}(R) = R^2/\nu$ to be less than the time scale for changes in the 
accretion rate $\dot M_{acc}/\ddot M_{acc}$. This yields a steady disk extent $R_s$. 

The disk temperature $T_d$ and sound speed $c_d$ are given by standard  expressions for an optically thick and isothermal structure \citep{1981ARA&A..19..137P}.
The accretion energy is radiated locally:
\begin{equation}
    T_d^4(R,t) = \frac{3GM(R,t)\dot M_a(t)}{8{\pi}{\sigma}R^3} \left[1\,-\,\left(\frac{R_*}{R}\right)^\frac{1}{2}\right],
\end{equation} 
where $M(R,t) = M_*(t)\,+\,M_d(R,t)$ is the sum of the protostellar and disk mass internal to $R$.
The sound speed for a molecular gas disk with a mean molecular weight of 2.3 is $c_d = 6.01\,10^3\,T_d^{1/2}$\,cm\,s$^{-1}$.

The disk mass $M_d$ is not expressible analytically since the accretion is driven by two viscosity components with differing functional forms. 
First, we take the usual turbulent viscosity $\nu_t = {\alpha_d}c_dH$ where H is the disk thickness and $\alpha_d$ is a dimensionless parameter which 
we set to 0.1 unless otherwise stated \citep{1973A&A....24..337S}. This yields
\begin{equation}
    \nu_t = \frac{2\,\alpha_d\,c_d^2}{3\,\Omega},
\end{equation}
where the angular rotation speed is $\Omega\,=\,{\surd}(GM/R^3)$. The  component of viscosity related to self-gravitational forces is 
parameterised, as suggested by \citet{1987MNRAS.225..607L}:
\begin{equation}
    \nu_g = \frac{2{\mu_d}c_d^2}{3\Omega}\left(\frac{Q_c^2}{Q_t^2} - 1\right)
\end{equation}
for $Q_t < Q_c$ and $\nu_g = 0$ otherwise. Here, the efficiency parameter  $\mu_d$ and the instability parameter $Q_c$ will also be set to unity
(see \citet{1990ApJ...358..515L}). Hence, the parameter
\begin{equation}
    Q_t = \frac{c_d\,\Omega}{\pi\,G\,\Sigma}
\end{equation} 
determines the importance of viscosity through the disk's own gravity. Note that the viscosity can be determined by self-gravity even when the 
protostellar mass is large when the disk column density is large. This  may well arise where the turbulent viscosity is inefficient, in the outer 
disk regions, leading to a build up of mass until self-gravity takes effect. 

The disk column density is given by 
\begin{equation}
    \Sigma = \frac{1}{2\pi\,R}\frac{dM}{dR} 
\end{equation}
and the cumulative mass distribution
\begin{equation}
    \frac{dM}{dR} = \frac{R\,\dot M_{acc}}{\nu_t\,+\,\nu_g}
\end{equation}
is given by the viscosity. From the disk column it is straightforward to calculate the disk radius at which the optical depth is unity, $R_{{\tau}=1}$.  
 This completes the set of equations which is integrated from an inner radius to yield the full disk structure. The results for disks associated  
low-mass protostars were presented  by \citet{2000IrAJ...27...25S} and will be extended in a following work.

\section{Results}            
\label{results}

\subsection{Clump Mass v. Bolometric Luminosity}
\label{clumps}

\begin{figure}
\includegraphics[width=8.7cm]{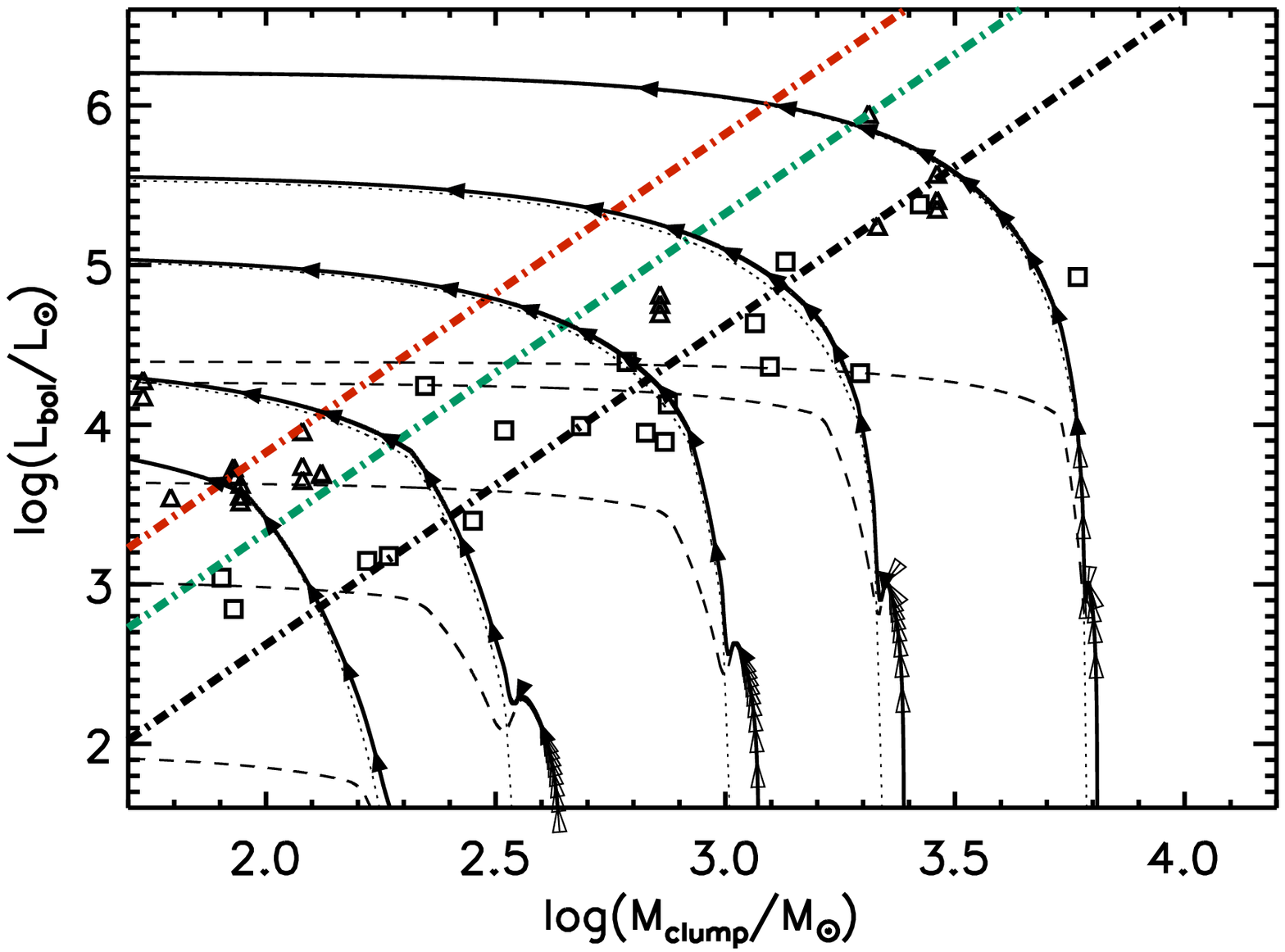}
\includegraphics[width=8.7cm]{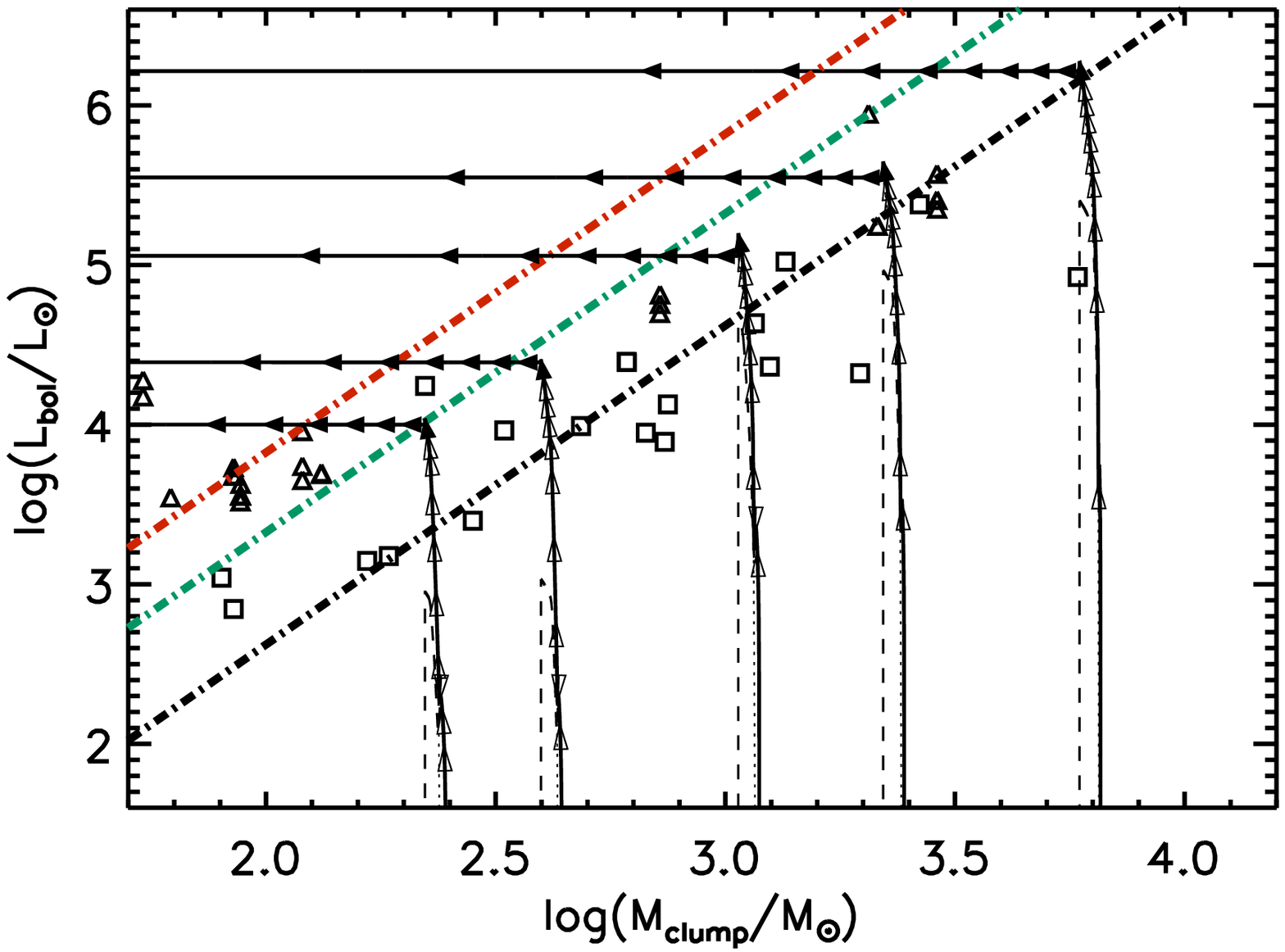}
\caption{ L$_{bol}$--M$_{clump}$ diagrams for Constant Slow Accretion (simultaneous star and clump evolution, upper panel) and 
Constant Fast Accretion (early formation of the massive star, lower panel), both forming via hot accretion.. 
The five solid lines correspond to  systematically increasing final stellar masses of 10, 15, 30, 50 and 100\,M$_\odot$. 
Initial clump masses, final stellar mass, accretion timescale and accretion rates are provided in Table~\ref{maintable}.
The contributing luminosity components derived from accretion (dashed lines) and interior (dotted lines, where visible)  are indicated.   The filled arrowheads are laid at equal time intervals of one tenth the total time, and the ten unfilled arrowheads are similarly placed  in the first of these intervals.
For comparison, we overlay (1) the data set presented by \citet{2008A&A...481..345M} where  the triangles denote the infrared sample while the squares denote the millimetre sample. 
The three straight dot-dashed lines correspond to bolometric temperatures of  40\,K (blue), 60\,K (green) and 80\,K (red)  as calculated by Method~2 with  an envelope density  radial power-law index of 1.5.  
 \label{lbol-menv0}}
\end{figure}
  
\begin{table*}
  \caption[table1]
   {Model data employed in Figs.~\ref{lbol-menv0} and \ref{lbol-menv1} in forming a 100~M$_\odot$ star from an initial clump of mass 6,561~M$_\odot$ (left) and  a 10~M$_\odot$ star from an initial clump of mass 247~M$_\odot$ (right), all  with 30\% of the mass ejected through jets or outflow.}
       \label{maintable}
  \begin{tabular}{l|lrr|lrr}
        \hline  \noalign{\smallskip}
     \sf{Accretion Model}            &  \sf{Max. accretion}  & \sf{Accretion} & \sf{Clump}
      &  \sf{Max. accretion }  & \sf{Accretion} & \sf{Clump }      \\
       & \sf{rate}  & \sf{timescale} &   \sf{timescale}
                                                      &  \sf{rate}  & \sf{timescale} &   \sf{timescale}    \\
                                                           
                                                      &  \sf{10$^{-3}$~M$_\odot$~yr$^{-1}$ }  & \sf{10$^{5}$~yr} &   \sf{10$^{5}$~yr}
                                                      &  \sf{10$^{-3}$~M$_\odot$~yr$^{-1}$ }  & \sf{10$^{5}$~yr} &   \sf{10$^{5}$~yr}    \\
     \noalign{\smallskip} \hline
     \noalign{\smallskip}
        {Constant - Slow }        &    0.141    &       10.0     &  10  &   0.0014  &   100  &  100\\
        {Constant - Fast }         &    1.413    &         1.0     &  10   &   0.0141  &    10   &   100 \\
         {Accelerated}   &    2.717    &         1.0     &  10   &   0.0907  &     3    &      10    \\
         {Power Law}    &     2.861      &         0.2     &  10   &   0.2861  &    0.2  &      10   \\
         {Exponential}   &     1.980      &         1.0     &   10  &   0.1980  &     1.0  &     10\\
              & & \\
        \noalign{\smallskip}  \hline  \noalign{\smallskip}
 \end{tabular}
\end{table*}

\begin{figure}
\includegraphics[width=8.7cm]{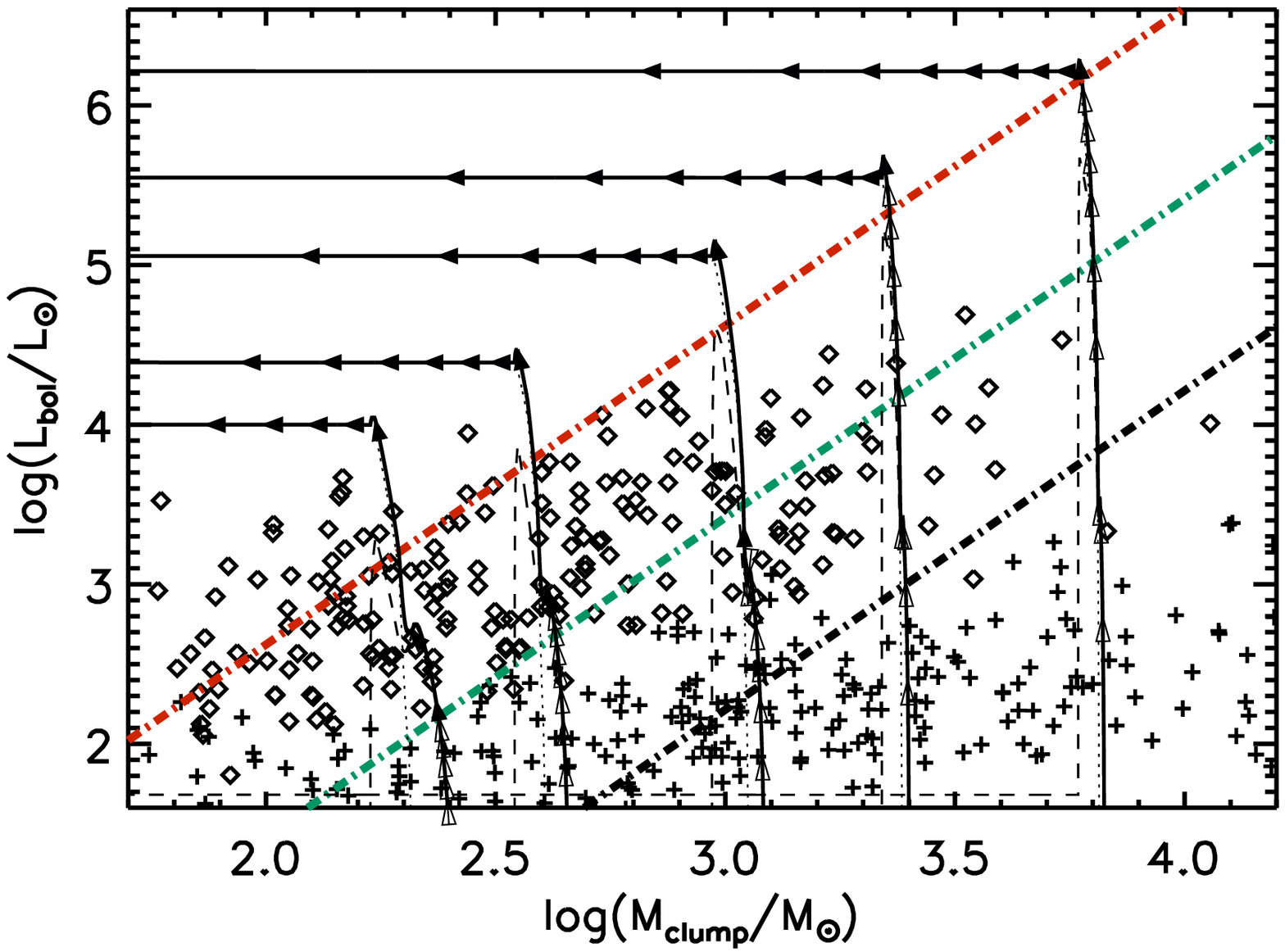}
\includegraphics[width=8.7cm]{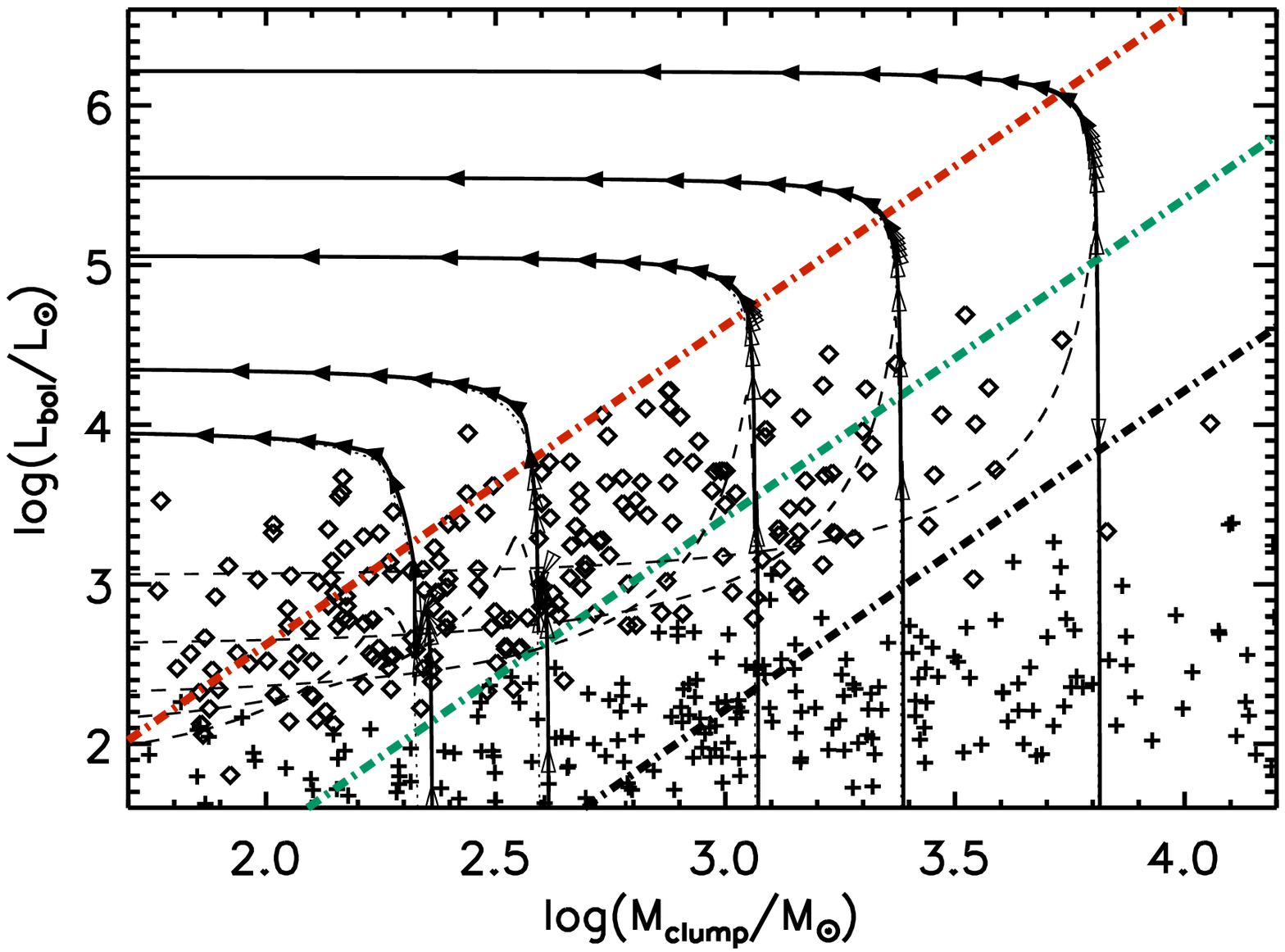}
\includegraphics[width=8.7cm]{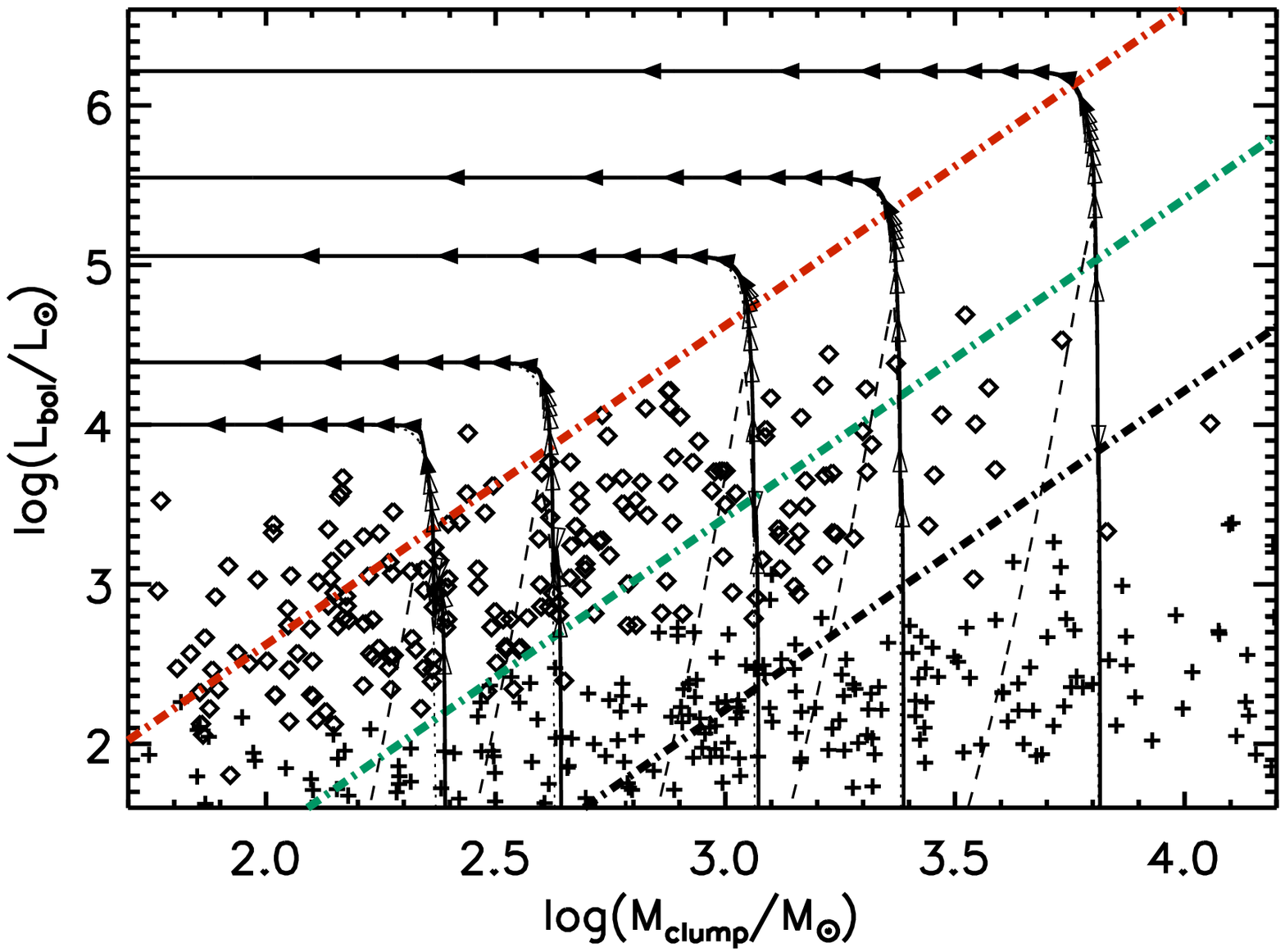}
\caption{ L$_{bol}$--M$_{clump}$ diagrams for Accelerated (upper), Power Law (middle) and Exponential (lower panel) accretion models assuming hot accretion.
The model parameters  are provided in Table~\ref{maintable}. The data are the set of Herschel objects  from the $l = 30^\circ$ field from the Herschel Infrared GALactic plane survey (Hi-Gal) \citep{2010A&A...518L..97E} as revised by \citet{2013A&A...549A.130V}. The three straight lines correspond to bolometric temperatures of  10\,K (blue), 20\,K (green) and 40\,K (red)   as calculated by Method~2 with  an envelope density  radial power-law index of 1.5. See the caption of Fig.~\ref{lbol-menv0} for other details.  \label{lbol-menv1}}
\end{figure}
 
 Cloud mass, bolometric luminosity and bolometric temperature are quantities derived from observations.  The bolometric temperature is sensitive to the geometry, orientation 
 and uniformity of the clump. Therefore, the  relationship  between cloud mass and bolometric luminosity  is most often employed as a diagnostic tool
 for the formation of high-mass stars \citep{2008A&A...481..345M,2013A&A...552A.123B} as well as lower-mass stars \citep{1993A&A...273..221R,1996A&A...309..827S,2000IrAJ...27...25S}.
 More recently, L$_{bol}$--M$_{clump}$ or L$_{bol}$/M$_{clump}$--M$_{clump}$ 
 diagrams have been utilised to analyse   Herschel Space Telescope data  \citep{2010A&A...518L..97E,2012A&A...547A..49R}.
 
 To interpret these data points as a time sequence, it is assumed that the clump mass decreases in time as the luminosity increases \citep{1993ApJ...406..122A}. 
Model evolutionary tracks are then easily calculated based on the principles discussed in Section~\ref{method} upon choosing an accretion type and rate.

The initial clump mass is assumed to be directly related to the final mass of the most massive star, $M_{*f}$, by the relation
\begin{equation}
      \log M_{clump}(0) = A  + 1.41 \log M_{*f}.
      \label{clumpmass}
\end{equation}
Here,  $A = 0.55$  could be taken \citep{2008A&A...481..345M} on the assumption that the clump generates the bound cluster \citep{2009A&A...503..801F}.  However, accounting for the global gas escape as well as that from protostellar outflows, we shall take $A = 0.85$  corresponding to a star formation efficiency of 50\%. As will be seen below,
such high efficiencies are inconsistent and much larger values of $A$ need to be considered. 

 The clump gas is taken to be reduced at  a constant rate (excluding the inner envelope which supplies the  massive star). However, other relationships between the stellar cluster 
and most-massive star have been considered and, above all,  there is a wide spread in the measured values of $A$  \citep{2010MNRAS.401..275W} corresponding to  an order 
of magnitude in mass. 

Diagrams for models with constant hot accretion  are shown in Fig.~\ref{lbol-menv0} and for variable rates in Fig.~\ref{lbol-menv1}. Also shown are observational data 
for far-infrared and infrared sources and the theoretical bolometric temperature isotherms. Note that these isotherms as calculated from Equation~\ref{tbolometric} are in agreement with the bolometric temperatures directly derived in the literature.

The top panel of Fig.~\ref{lbol-menv0} displays the tracks for  a slow  evolution \citep[e.g.][]{2001A&A...373..190B} with a simultaneous cloud evolution. This yields a strong dependence between the two parameters with a wide distribution  of sources predicted. In recent works, the clump mass is assumed to fall at a slow constant rate with time while the massive star  forms abruptly from a compact envelope, which is consistent with both available data and theoretical expectations  \citep{2002Natur.416...59M}. We therefore discount the Slow Accretion scenario as a model for
clump/cluster evolution although it is relevant in following the possible evolution of an isolated core especially in the low-mass case.

The constant Fast Accretion case is shown in the lower panel of of Fig.~\ref{lbol-menv0}. It is clear that massive infrared protostars would be rarer in this  case (as indicated by the arrowheads placed at regular time intervals). In addition, a much narrower range in clump masses for a given luminosity interval is predicted in the latter case.

The Accelerated Accretion Model generates similar results (top panel of Fig.~\ref{lbol-menv1}) with two distinct track stages. This was the model explored by  \citet{2008A&A...481..345M} but we note here that both the constant and power law fall-off models produce very similar tracks. To differentiate between models will require a detailed statistical  analysis. 
The Power Law Model generates tracks in which there is an extended transition stage between the accretion and clean-up stages (middle panel of Fig.~\ref{lbol-menv1}). This transition  stage
occurs at bolometric temperatures between 60\,K and 80\,K. 

\begin{figure}
\includegraphics[width=8.7cm]{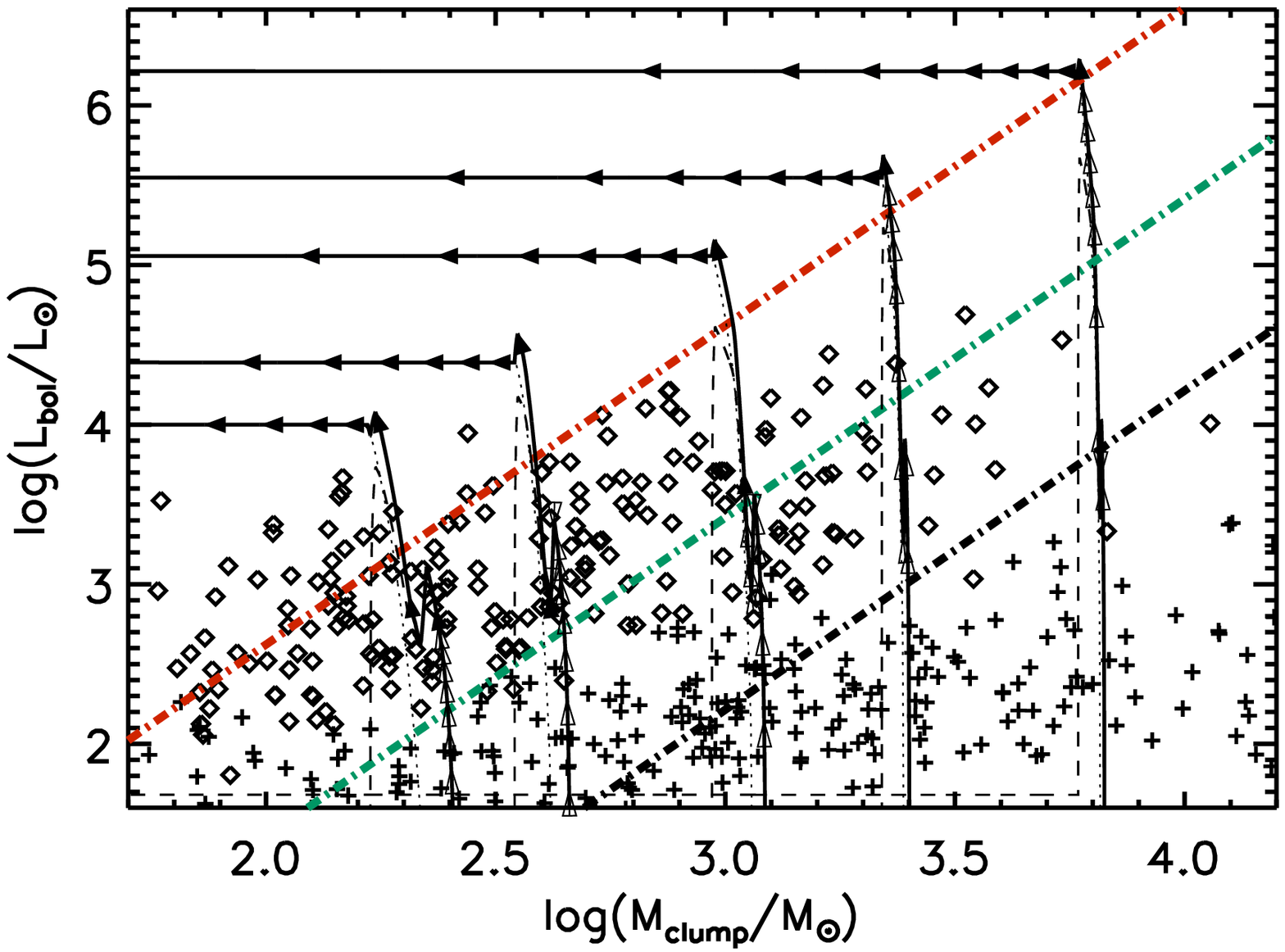}
\includegraphics[width=8.7cm]{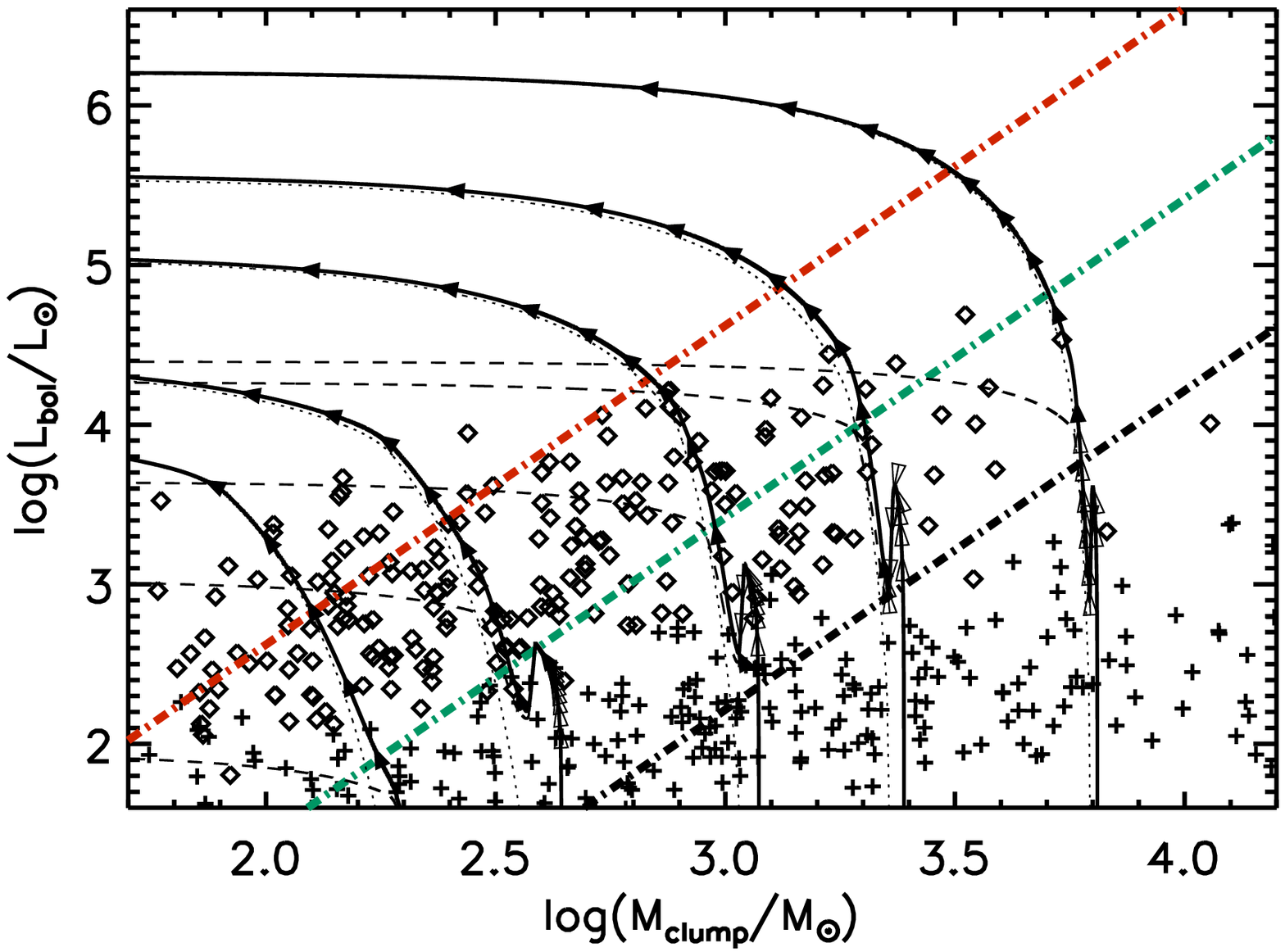}\caption{ Cold Accretion: L$_{bol}$--M$_{clump}$ diagrams for the Accelerated Accretion Model and the Constant-Slow Accretion Model with Cold Accretion. The model parameters  are as on the corresponding Hot Accretion case provided in Table~\ref{maintable}. See the caption of Fig.~\ref{lbol-menv1} for further details.
  \label{lbol-menv2}}
\end{figure}

It should also be remarked that the track direction changes from almost vertically up to down for a short period. This is caused by the swelling of the protostar
which reduces the bolometric luminosity by reducing the accretion luminosity. While hardly visible in  Fig.~\ref{lbol-menv1}, this effect becomes prominent in the cold accretion scenario
where the expansion phase is stronger. This is shown in Fig.~\ref{lbol-menv2} for the Power Law Model in which it is most obvious. In general, however, we find that Cold Accretion does not significantly alter the tracks on the L$_{bol}$--M$_{clump}$ plots.

\subsection{Bolometric Temperature}

The bolometric isotherms calculated through Method 2 are shown on the  L$_{bol}$--M$_{clump}$ plots. The temperatures   are broadly consistent with the range of temperatures
derived from the spectral energy distributions for the Herschel sources. We do not expect more than this given the known sensitivity to
geometry and orientation. Indeed, as shown in Fig.~\ref{isotherms}, there is a strong dependence on the radial density distribution as given by the index $\beta$.

\begin{figure}
\includegraphics[width=8.7cm]{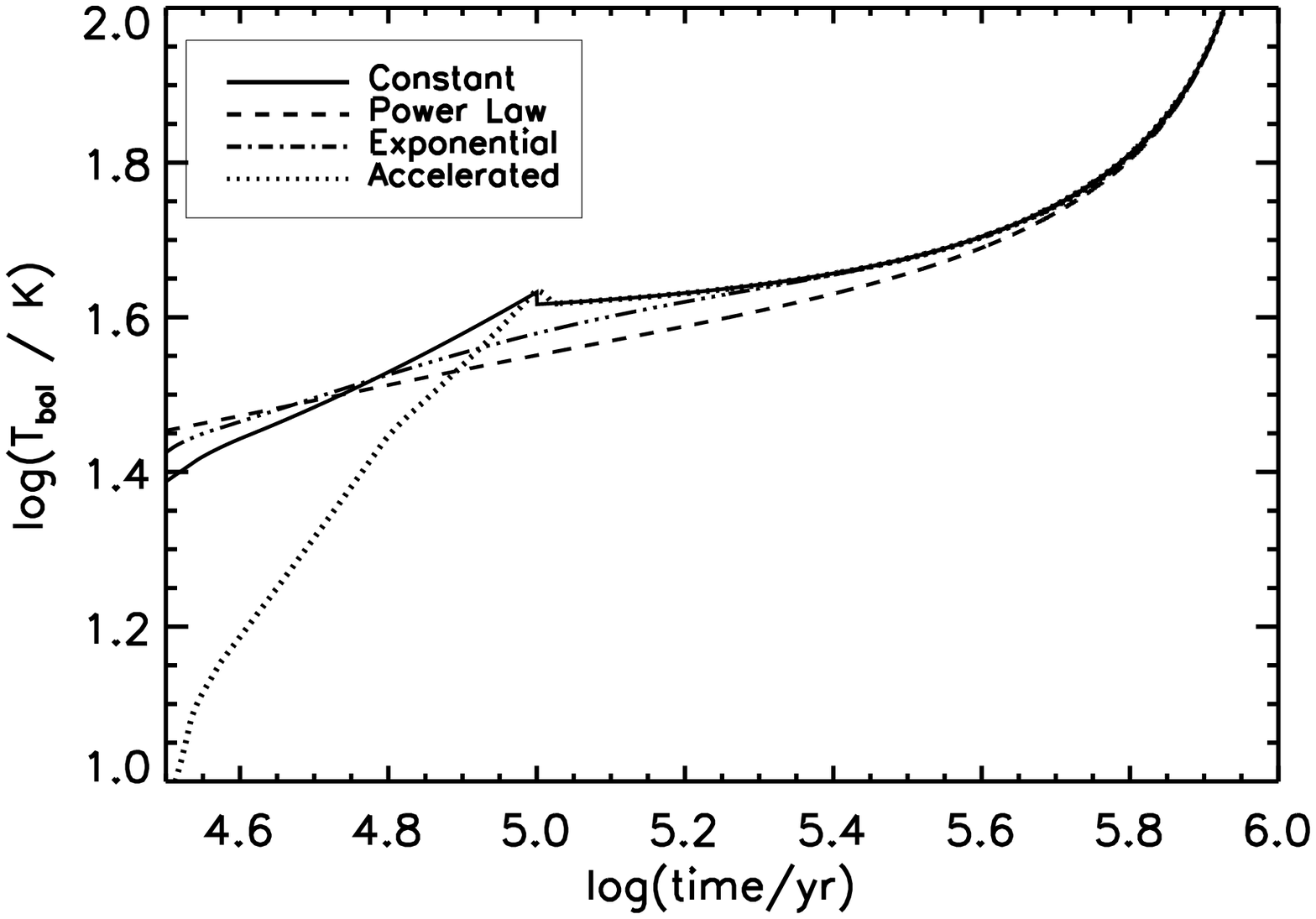}
\includegraphics[width=8.7cm]{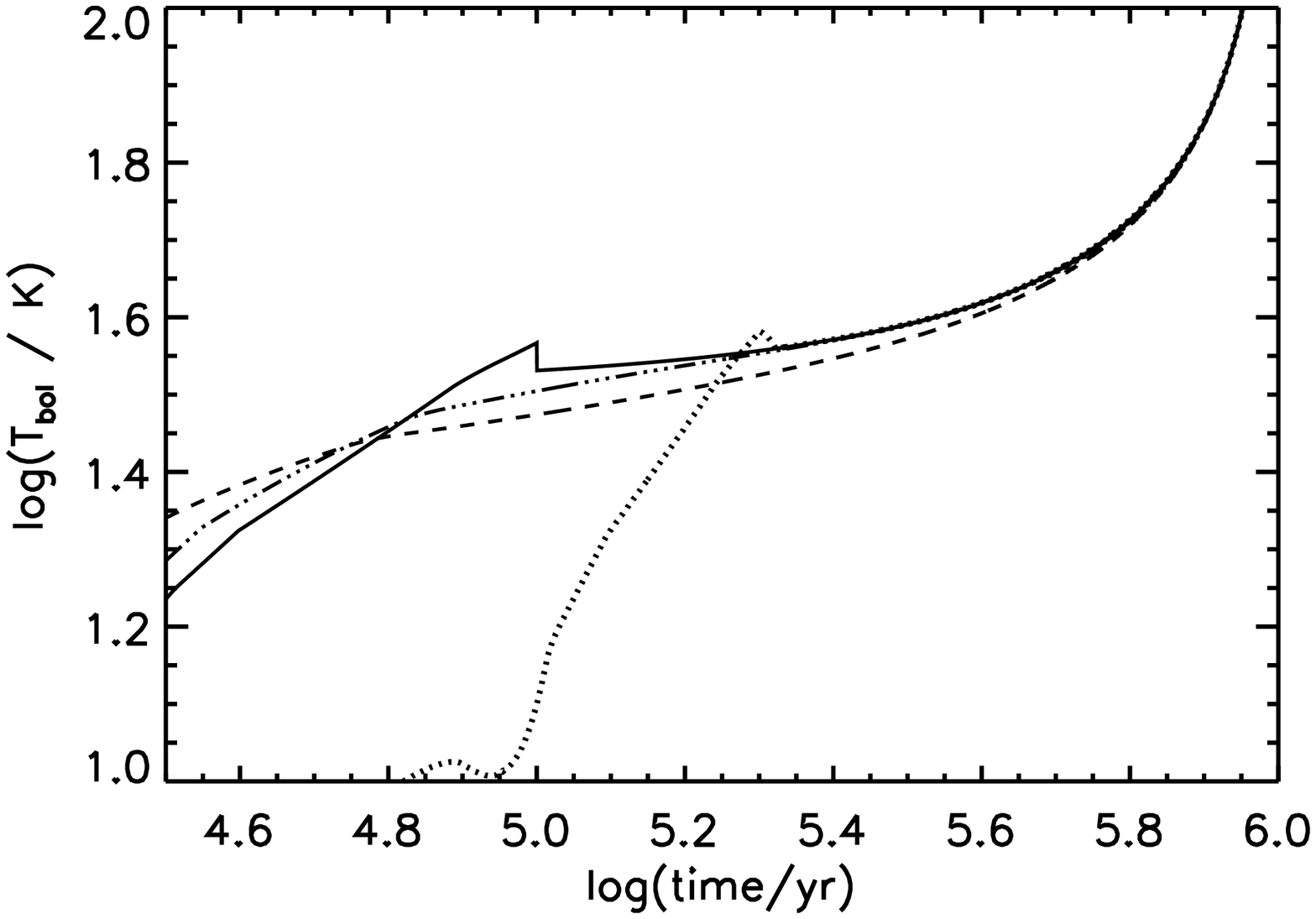}
\caption{The early evolution of the bolometric temperature for  Hot Accretion models  which yield  100~M$_{\odot}$ (upper panel) and 30~M$_{\odot}$ (lower panel) stars, with lines as denoted in Fig.~\ref{rateversustime} and parameters from Table~\ref{maintable}. Hot accretion is assumed here. The  panels assume that the initial clump mass is twice that necessary to generate the associated cluster ( A = 0.85).
 \label{tbolversustime}}
\end{figure}

However, we can compare the accretion models statistically to determine if they would yield significantly different statistics in terms of numbers of source in any temperature interval. These numbers are provided in Table~\ref{tabletime}.   As also  illustrated in Fig.~\ref{tbolversustime},  there is a remarkable difference between the Accelerated Accretion model and the alternative
evolutions. In the Accelerated Accretion Model, significantly more time elapses in the low temperature ($<$ 30\,K)  regime. For the 100\,M$_\odot$ case, all the other models investigated (with the exception of Slow Accretion) spend much less time at low temperatures, by a factor of two to  three.  This result is consistent with expectations: accelerated accretion takes time to get off the ground.
 
\begin{figure}
\includegraphics[width=8.7cm]{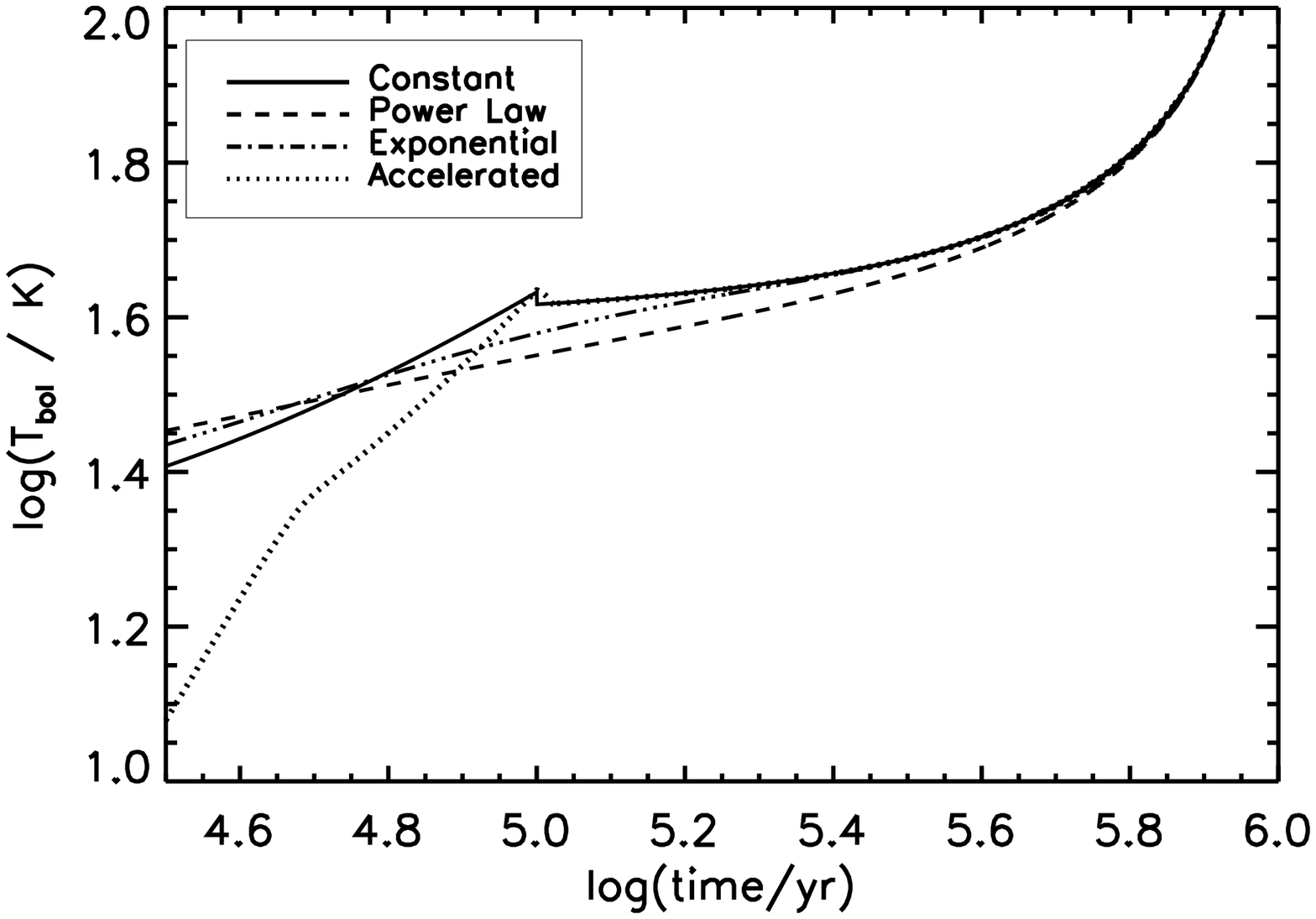}
\includegraphics[width=8.7cm]{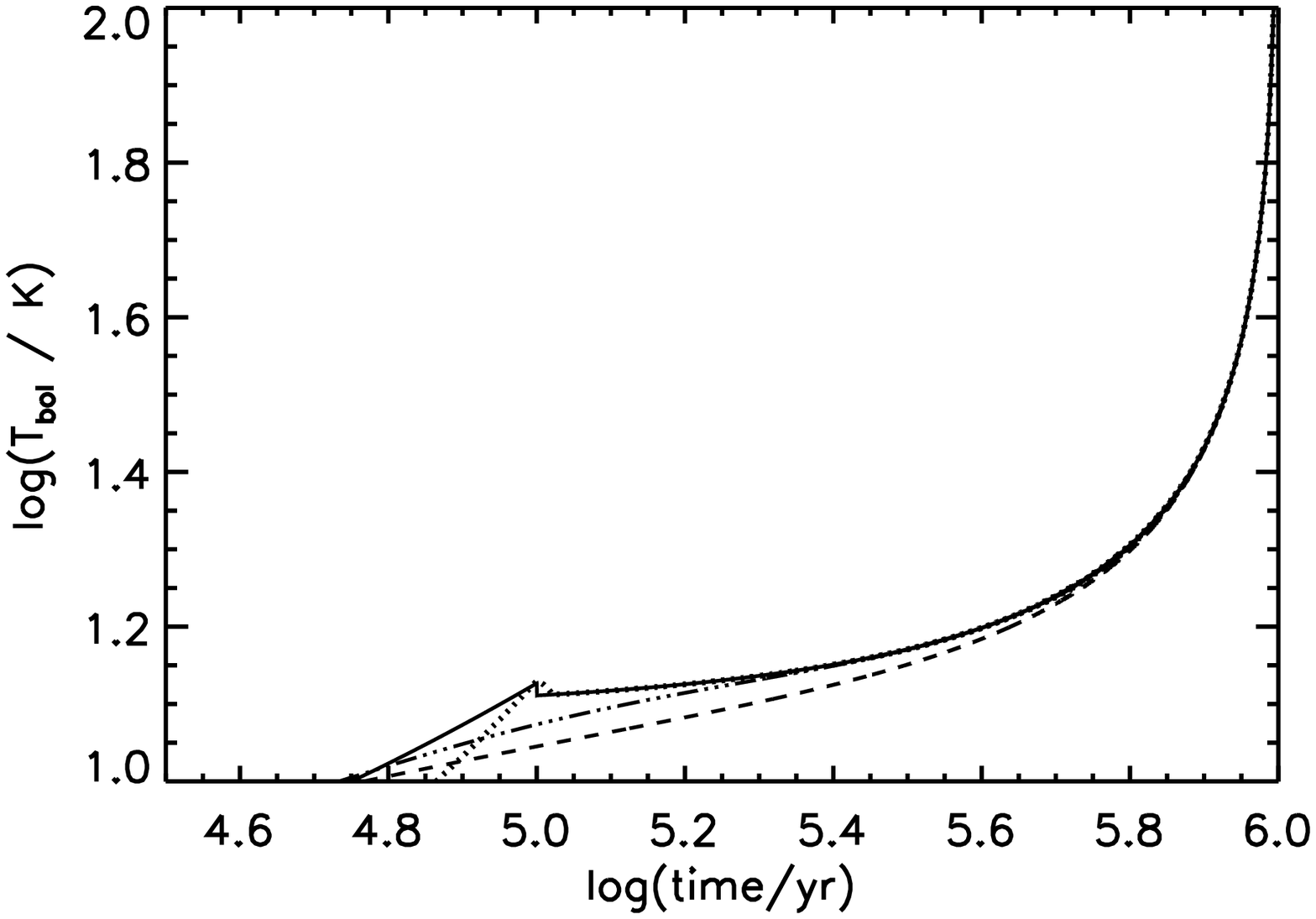}\caption{The early evolution of the bolometric temperature for  Cold Accretion  models  with 
 the initial clump mass twice (upper panel) and 9.4 times (lower panel)  the fiducial value required to generate the stellar cluster.
each model displayed forms  a 100~M$_{\odot}$ stars, with lines as denoted in Fig.~\ref{rateversustime} and parameters from Table~\ref{maintable}. 
Although cold accretion is assumed, hot accretion is found to closely   follow this behaviour. \label{tbolversustime2}}
\end{figure} 

\begin{table*}
  \caption[table1]
   {The fraction of the total time required to reach a bolometric temperature of 100\,K in order to traverse the indicated temperature range for the 
   models to form a star of mass 100~M$_\odot$ through hot (upper section) and cold (lower section) accretion. The clump mass is given by Equation~\ref{clumpmass}
   and the radius of the clump is 50,000~AU, consistent with the R$_{env}$ values derived by  \citet{2008A&A...481..345M}. Herschel Hi-Gal fractions are
   taken from the two fields analysed by \citet{2010A&A...518L..97E} and \citet{2013A&A...549A.130V}. }
       \label{tabletime}
  \begin{tabular}{l|rrrrrr}
        \hline  \noalign{\smallskip}
     \sf{Accretion Model}            &  \sf{$<20\,$K}  & \sf{ $<30\,$K } & \sf{30\,K -- 50\,K}
      &  \sf{50\,K -- 70\,K}  & ~~~~~\sf{$< 100$\,K time ($\times$ 1000 yrs)}     \\
     \noalign{\smallskip} \hline
     \noalign{\smallskip}
          & \\
          HOT ACCRETION: & \\
        {Constant - Slow}     &        0.250 &      0.442 &      0.279 &      0.158 &       866 \\ 
         {Constant - Fast}     &      0.027 &     0.057 &      0.397 &      0.356 &       845 \\
        {Accelerated}            &      0.058 &     0.080 &      0.373 &      0.356 &       845\\   
         {Power Law}          &    0.020 &     0.049 &      0.446 &      0.321 &       847\\ 
         {Exponential}         & 0.025 &     0.051 &      0.402 &      0.356 &       845 \\
         & & \\
         COLD ACCRETION: \\
         {Constant  - Slow}        &   0.250 &      0.442 &      0.279 &      0.158 &       866 \\   
      {Constant - Fast}              &    0.022 &     0.057 &      0.397 &      0.356 &       845 \\
                 {Accelerated}   &  0.052 &     0.080 &      0.373 &      0.356 &       845\\         
                  {Power Law}    &  0.017 &     0.049 &      0.446 &      0.321              847\\     
    {Exponential}   & 0.021 &     0.051 &      0.402 &      0.356 &       8459 \\         
        & & \\
        Hi-Gal Data: (Elia et al 2010) &                 &                  &                &     & number    \\
$l = 30^\circ$ field &   0.463           &     0.832  &     0.128  &    0.022  &      311   \\   
$l = 59^\circ$ field & 0.417            &     0.846   &    0.142   &    0.011 &     91       \\
H-Gal YSOs (Veneziani et al 2013)  &  0.463   &     0.989  &     0.011   &   0   &      284  \\
        \noalign{\smallskip}  \hline  \noalign{\smallskip}
 \end{tabular}
\end{table*}
 
\begin{table*}
  \caption[table2]
   {The fraction of the total time required to reach a bolometric temperature of 100\,K in order to traverse the indicated temperature range for cold accretion 
   with fixed clump mass as indicated but different final stellar masses. The radius of the clump is 30,000~AU 
 consistent with the  values derived by \citet{2013A&A...549A.130V}. }
       \label{tabletime2}
  \begin{tabular}{l|rrrrrr}
        \hline  \noalign{\smallskip}
     \sf{Final Stellar Mass}            &  \sf{$<20\,$K}  & \sf{ $<30\,$K } & \sf{30\,K -- 50\,K}
      &  \sf{50\,K -- 70\,K}  & \sf{$< 100$\,K time (yrs)}     \\
     \noalign{\smallskip} \hline
     \noalign{\smallskip}
          & \\
          {Constant - Fast}         & \\
                    CLUMP MASS:  6,683~M$_\odot$ & \\
                    CLUMP RADIUS:  50,000~AU & \\
                 100  & 0.022 &     0.057 &      0.396 &      0.356 &       845\\ 
                    50  &  0.042 &      0.233 &      0.538 &      0.148 &       929 \\
                    30  &  0.077 &      0.581 &      0.294 &     0.081 &       960\\
                    15   &  0.550 &      0.810 &      0.133 &     0.036 &   1472 \\
                     10  &  0.715 &      0.880 &     0.084 &     0.023 &   988\\           
        & & \\
          CLUMP MASS: ~5,740~M$_\odot$ & \\
            CLUMP RADIUS: ~30,000~AU & \\
           STELLAR MASS:~30~M$_\odot$ \\ 
                 {Constant  - Slow}   &   0.672 &      0.826 &      0.114 &     0.037 &   2,936 \\                  
      {Constant - Fast}     & 0.473 &      0.778 &      0.156 &     0.043 &   2,935 \\       
          {Accelerated}   & 0.473 &      0.777 &      0.156 &     0.043 &       978 \\     
            {Power Law}    & 0.497 &      0.781 &      0.153&     0.042 &       978\\ 
    {Exponential}   &0.472 &      0.777 &      0.156 &     0.043 &       978 \\ 
                  & \\        
                  Hi-Gal Data: &                 &                  &                &  &    number    \\
        $l = 30^\circ$ field    &   0.463           &     0.832  &     0.128  &    0.022  &      311   \\   
        $l = 59^\circ$ field                        & 0.417    &     0.846   &    0.142   &    0.011 &     91       \\  
H-Gal YSOs (Veneziani et al 2013)  &  0.463   &     0.989  &     0.011   &   0   &      284  \\
        \noalign{\smallskip}  \hline  \noalign{\smallskip}
 \end{tabular}
\end{table*}

\subsection{Herschel Hi-Gal data}
 \label{herschel}
 
  Data directly comparable to that predicted in Table \ref{tabletime} are available through several Herschel programmes. The distribution of bolometric temperatures in Table~\ref{tabletime} along with the
data employed here are inconsistent. Over  80\% of observed clumps have temperatures below 30\,K  (bottom lines in Table~\ref{tabletime}) whereas the models predict a much more even number distribution with temperature. It is clear that only a small fraction of the Herschel cores on the very high mass tracks  will go on to form such massive stars. 

\begin{figure}
\includegraphics[width=8.7cm]{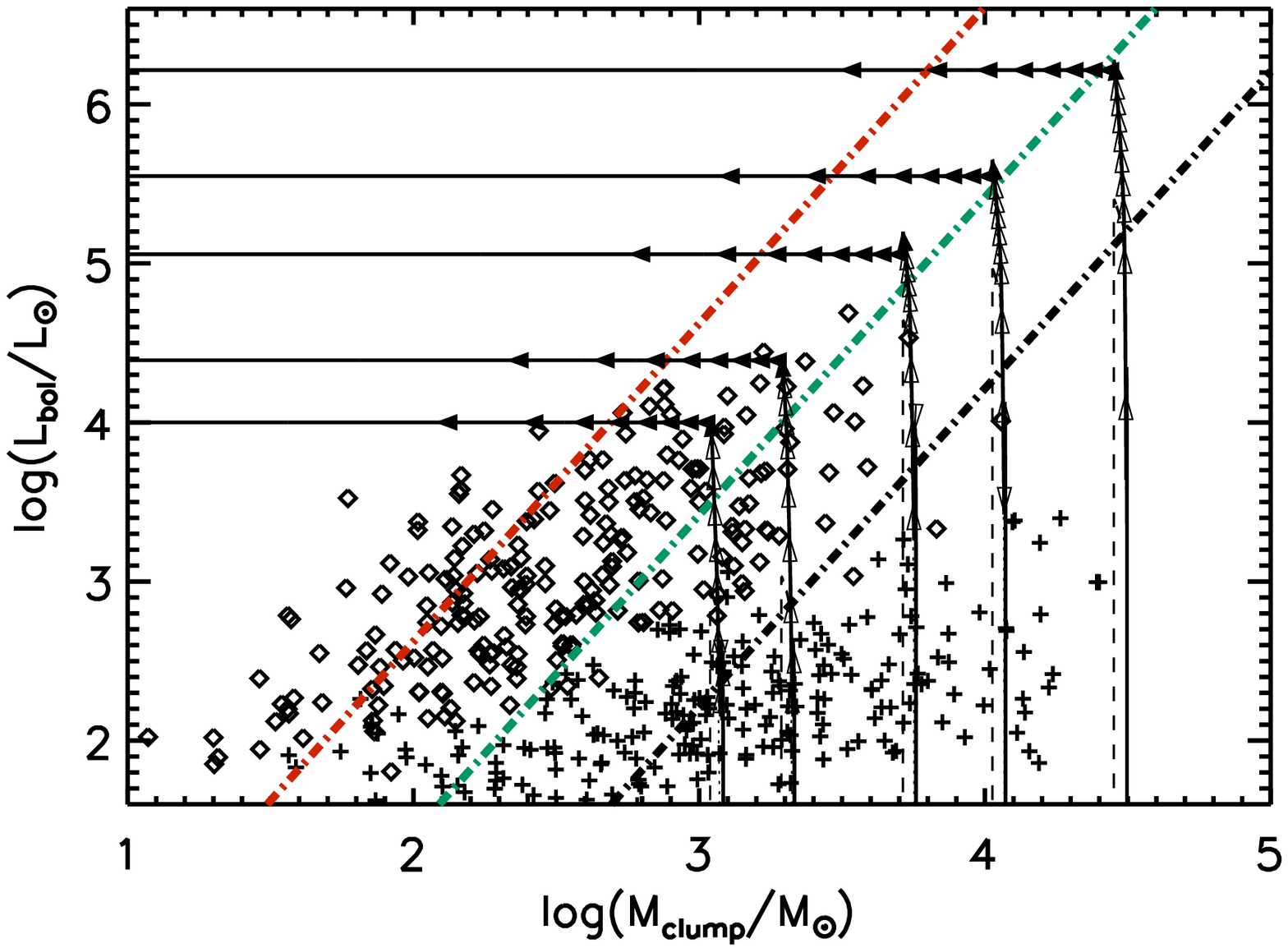}
\caption{Revised  L$_{bol}$--M$_{clump}$ tracks for the Constant Accretion model with
fast cold accretion . 
The six solid lines correspond to  systematically increasing final stellar masses of 8, 10, 15, 30, 50 and 100\,M$_\odot$. 
The Constant Accretion Model is taken here with final stellar mass, accretion timescale and accretion rates provided in Table~\ref{maintable} but the initial clump has
been increased by a factor of 4.7 corresponding to an initial clump mass 9.4 times larger than the final cluster mass.
The data are from the Herschel $l = 30^\circ$ field  as analysed  by \citet{2013A&A...549A.130V} with protostellar clumps (triangles) and pre-stellar clumps (crosses; 
taken as bound starless clumps if the temperature is used to derive the relevant internal measure of  pressure).\label{newtracks}  }
\end{figure}

 A resolution to this problem is straightforward:
the most massive forming star observed in the clumps is a factor of  about 3 smaller than that used in the literature to calculate tracks. 
In Table~\ref{tabletime2}, we present re-calculated 
number distributions on the assumption that the clump masses are indeed large but are actually being heated by  protostars which will form much lower mass stars. The new tracks are illustrated in Fig.~\ref{newtracks}. Extremely good correspondences are apparent.  Note that the initial clump masses are now 4.7 times larger, corresponding to stars of three times the mass according to Equation~\ref{clumpmass}.

This interpretation of the statistics, in which only 10--15\% of the initial clump mass ends up in stars, is independent of the accretion model. As shown on the panels of
Fig.~\ref{tbolversustime2}, it is very difficult to distinguish between the models for the high mass clumps.

The new interpretation is consistent with the data relating star clusters to the most massive stars as  presented by \citet{2010MNRAS.401..275W}.
Their Fig.~3 shows that there is a minimum mass of the most-massive star for a star cluster of a given size which is approximately three times lower than the average stellar mass. 
This minimum mass would, of course, be the most likely if drawn randomly from a distribution corresponding to the  Initial mass Function. 
We thus recommend that far-infrared data be interpreted by tracks as shown in Fig.~\ref{newtracks}.

 Fig.~\ref{newtracks} also demonstrates that the evolutionary phase of the observed sample of Herschel protostars is more advanced than previously interpreted
because the final mass of the stars had been overestimated. The assertion here is that there  are far fewer embedded protostars with mass exceeding 30~M$_\odot$.

\subsection{Radiative Feedback: hot accretion}

\begin{figure}
\includegraphics[width=8.7cm]{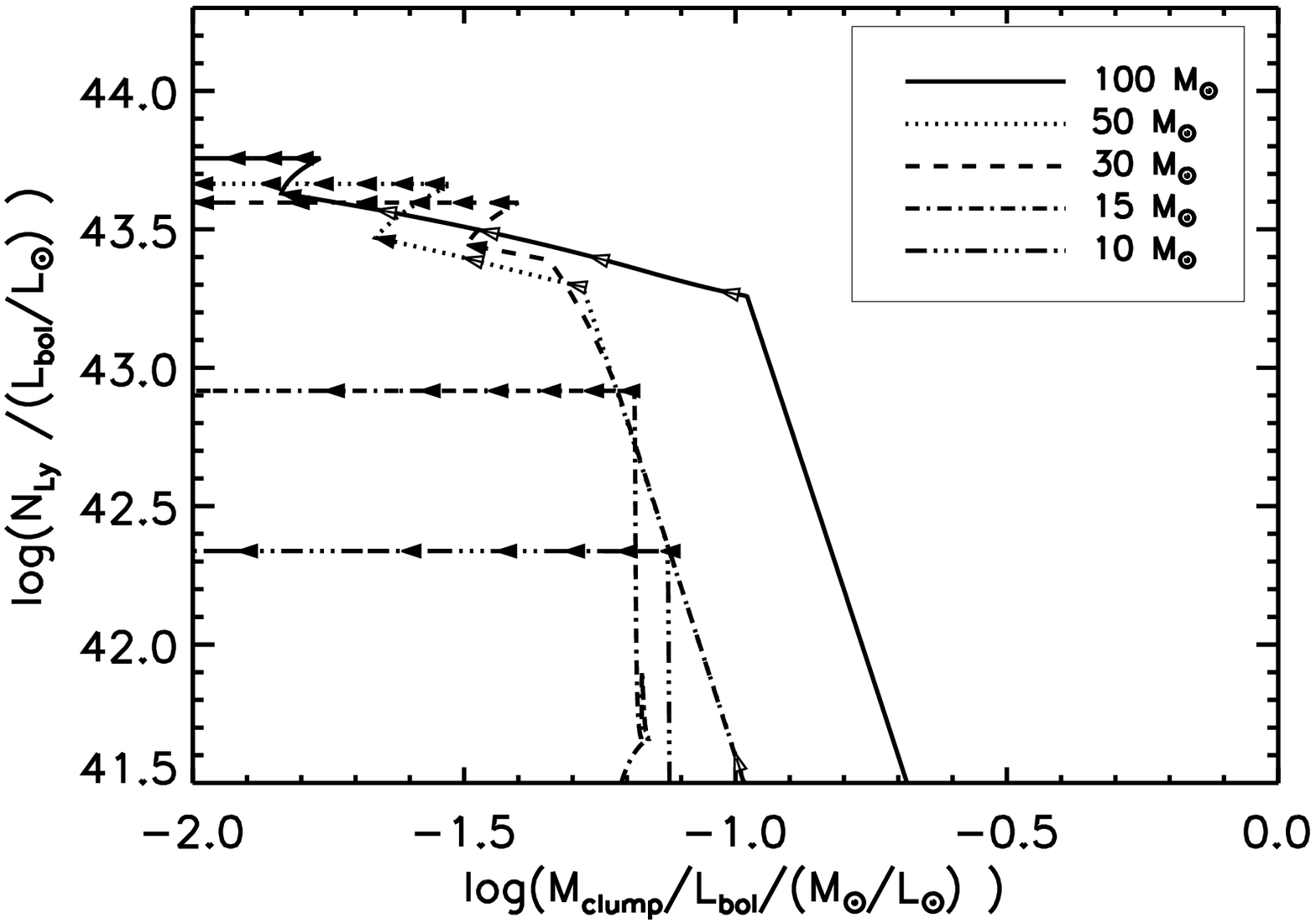}
\includegraphics[width=8.7cm]{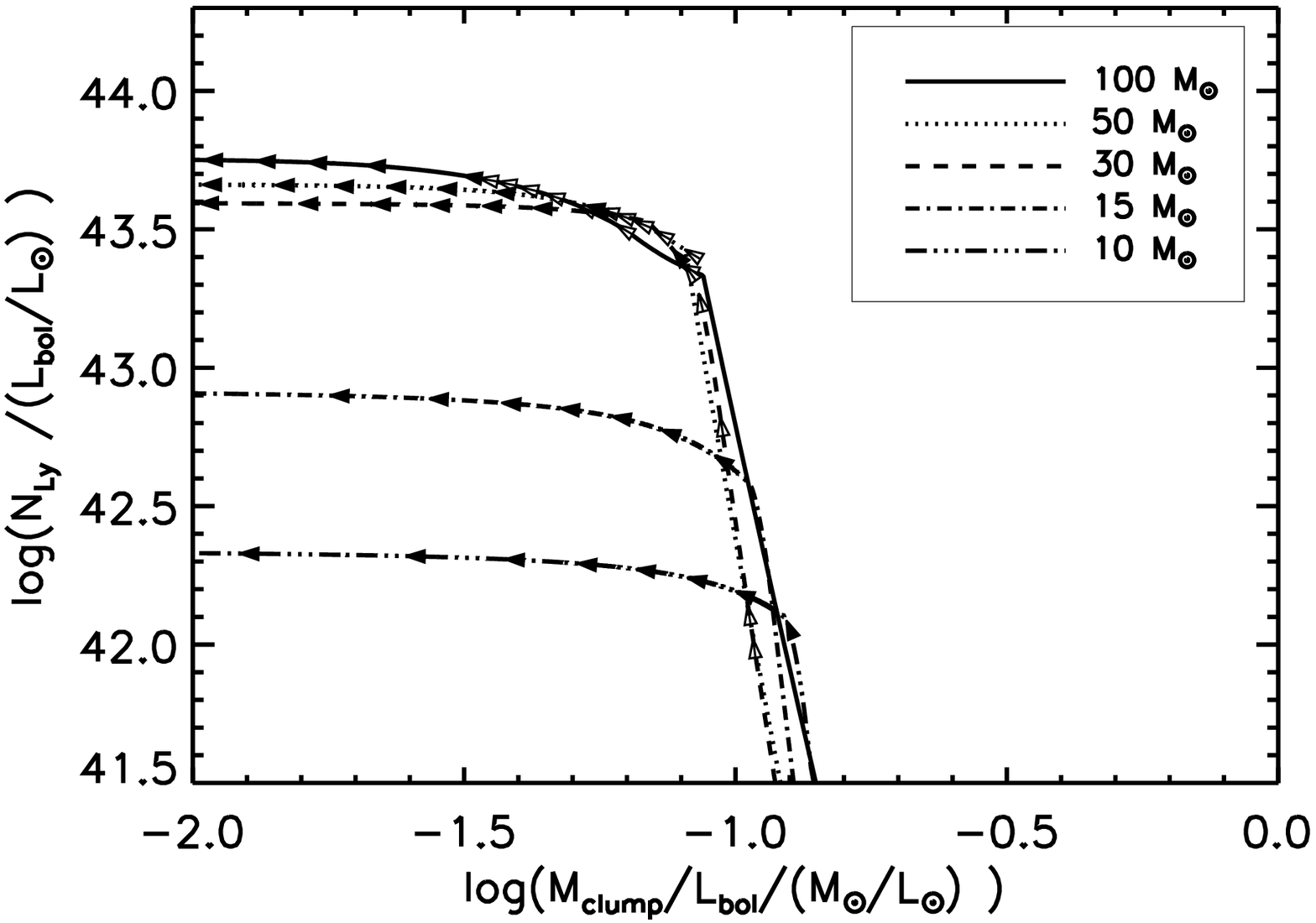}
\caption{The ratio of Lyman photon flux to bolometric luminosity, $N_{Ly}$/$L_{bol}$, against the evolutionary measure $M_{clump}/L_{bol}$. 
In this diagram, any large distance ambiguity is excluded. The Accelerating Accretion Model (upper panel) and Power Law Model (lower panel) tracks are displayed
for the Hot Accretion model. We also now implicitly assume the 4.7 times higher clump mass deduced in Sub-section \ref{herschel}. Here,  and in the following figures, the model tracks correspond to stars of final mass 100 (full) , 50 (dotted), 30 (dashed) , 15  (dot-dashed) and 10 M$_\odot$ (3-dot-dashed). \label{lyman5}}
\end{figure}

As the  protostar matures, the rapid increase in luminosity leads to a high surface temperature and a high number of extreme ultraviolet photons,  $N_{Ly}$, capable of generating 
a  surrounding source of free-free radio emission as quantified in Sub-section~\ref{radiation}. With this interpretation, we can compare two distinct indicators of evolution:
$M_{clump}/L_{bol}$ and  $N_{Ly}$/$L_{bol}$. Remarkably, both these quantities are, at least in principle,  distance independent. It is apparent from
Fig.~\ref{lyman5} that there is a significant difference between the two extreme models with the Accelerated Accretion tracks for the most massive stars occupying a wider region whereas the Power Law Model predicts much less variation.

\begin{figure}
\includegraphics[width=8.7cm]{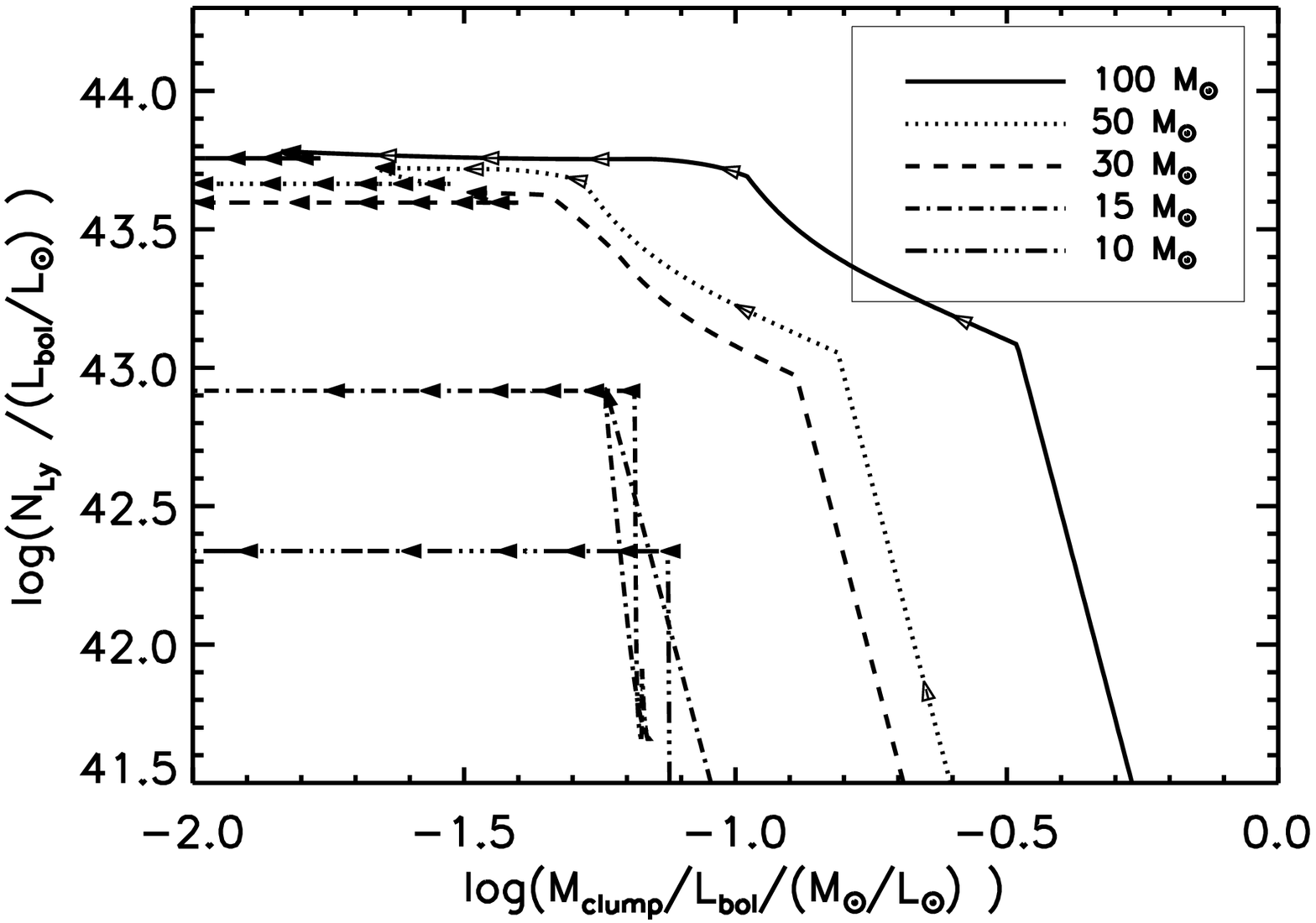}
\includegraphics[width=8.7cm]{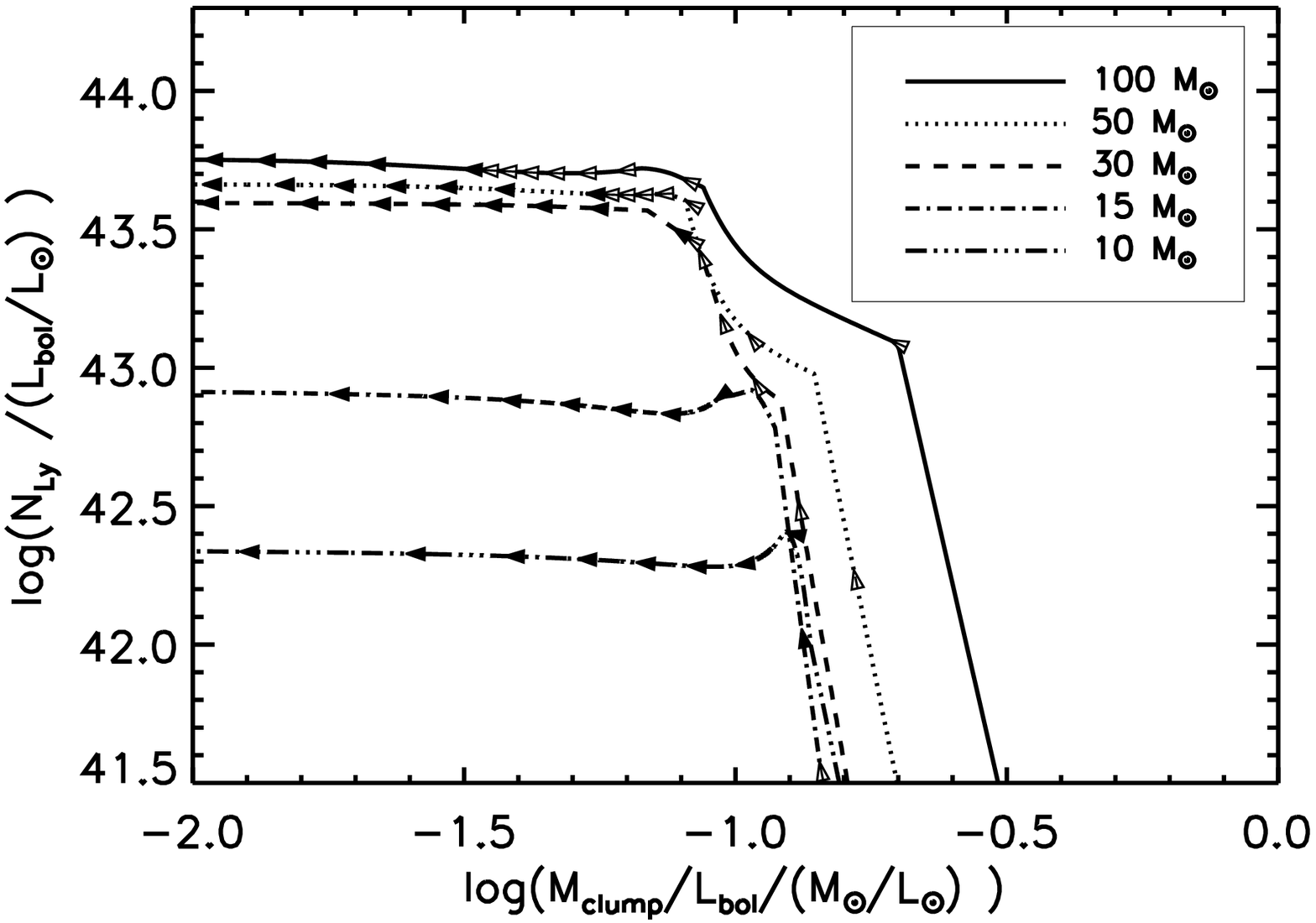}
\caption{Hot Spot Accretion. The ratio of Lyman photon flux to bolometric luminosity, $N_{Ly}$/$L_{bol}$, against the evolutionary measure $M_{clump}/L_{bol}$. 
A hot spot covering 5\% of the surface upon which 75\% of the accretion luminosity is emitted is assumed here. 
In this diagram, any large distance ambiguity is excluded. The Accelerating Accretion Model (upper panel) and Power Law Model (lower panel) tracks are displayed..  \label{hotspots}}
\end{figure}

We can now suppose that the inner accretion is not spherical but streams onto a limited area of the star, forming accretion hot spots on the surface. 
Taking a fraction $f_{acc}$ of the accretion to free-fall on to a fraction $f_{hot}$ of the surface area, yields a hot spot temperature $T_{hot}$ given by
\begin{equation}
     T_{hot}^4(R,t) = \frac{L_{int} + f_{acc} L_{acc}/f_{hot}  } {4{\pi}{\sigma}R_*^2}.
\end{equation} 
Contributions to the Lyman flux from the hot spot and the rest of the surface are then added. As shown in Fig.\ref{hotspots},
the behaviour is  different with the  the accretion luminosity generating significant early ultraviolet emission. Moreover, the hotspots are very important Lyman emittors for the stars of mass  10 -- 20~M$_\odot$ and can dominate the radio emission during the early phases of star formation.


\begin{figure}
\includegraphics[width=8.7cm]{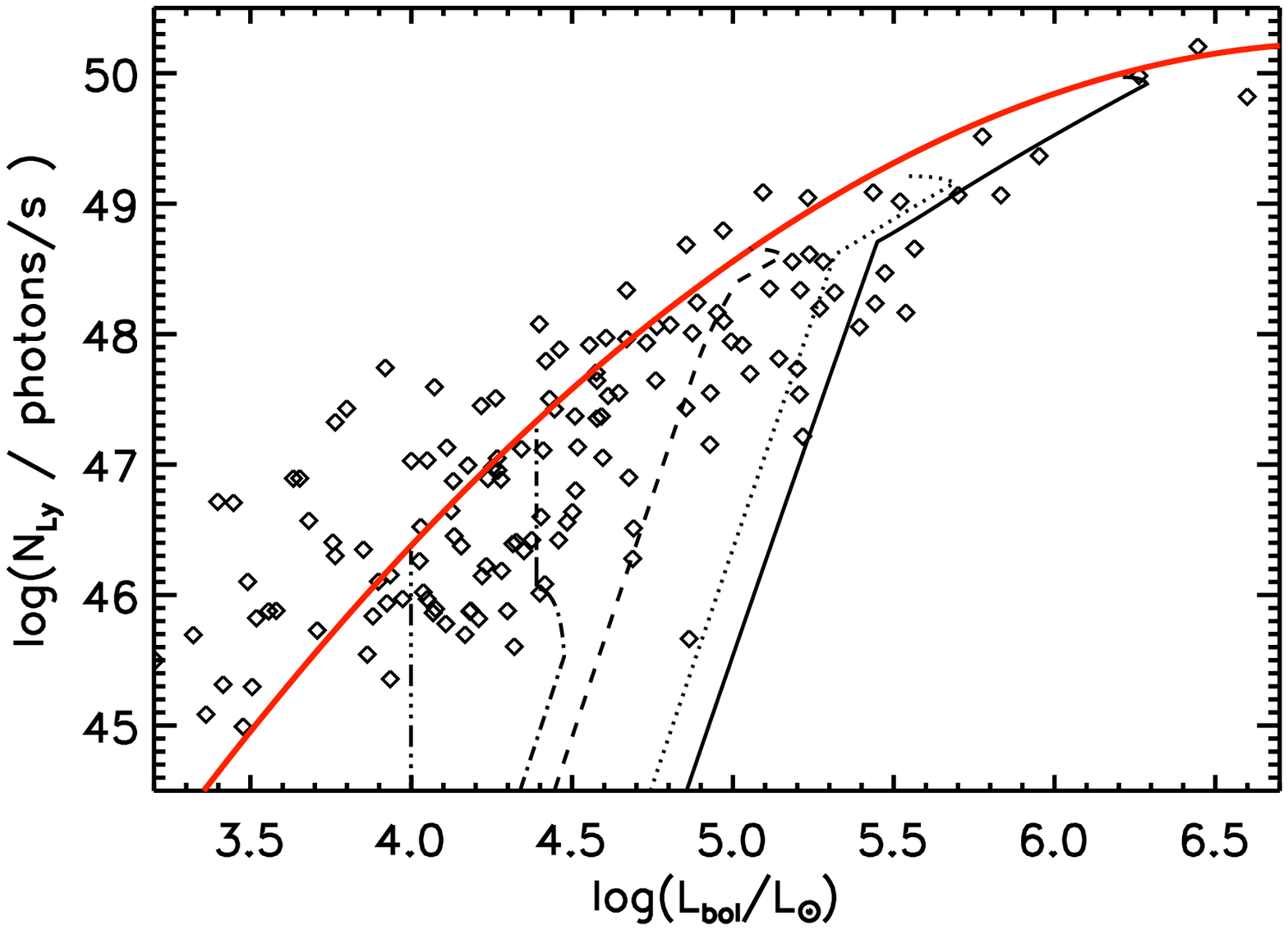}
\includegraphics[width=8.7cm]{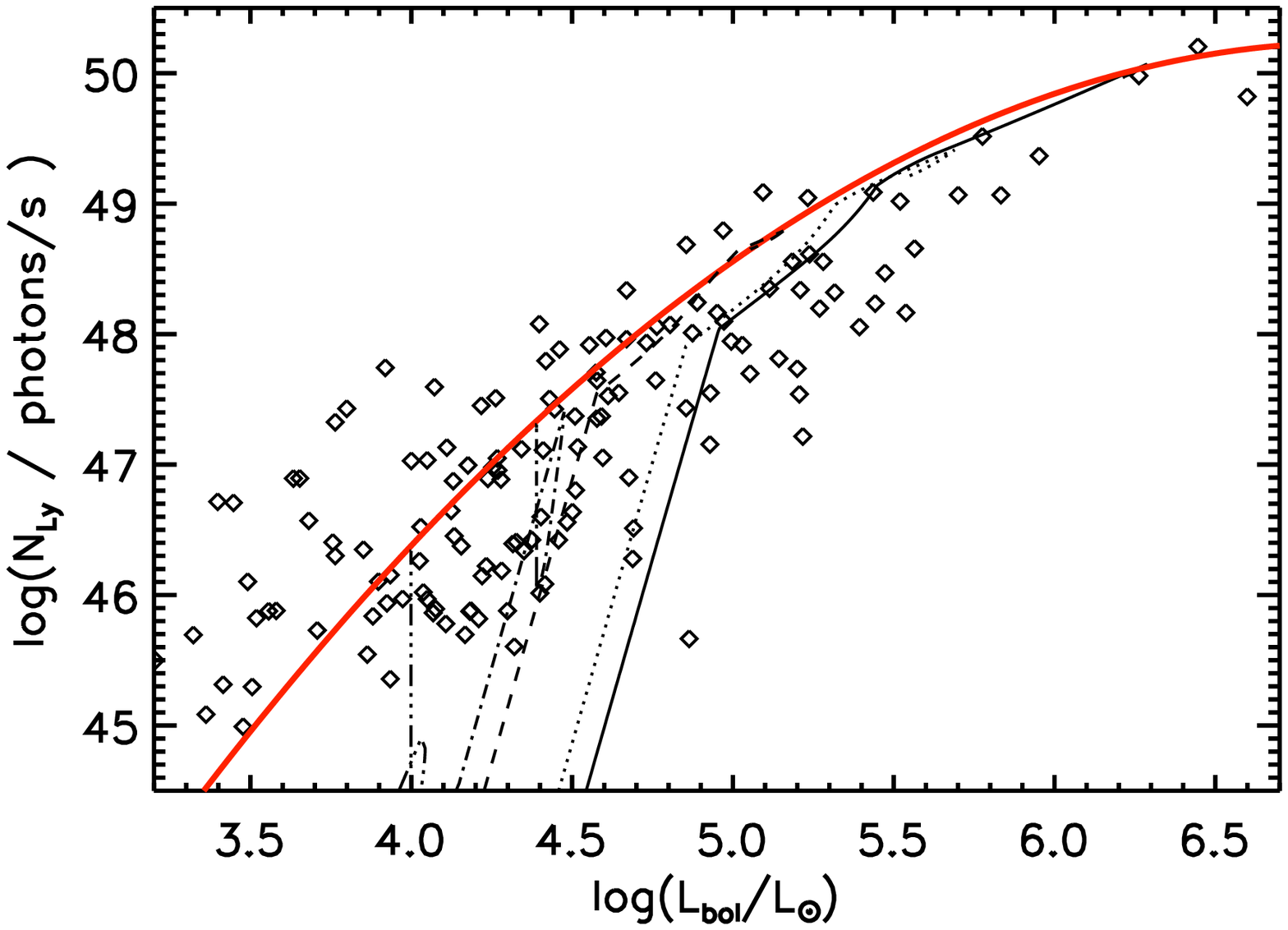}
\caption{The  Lyman photon flux, $N_{Ly}$, against  $L_{bol}$ for the Accelerated Accretion Model with hot accretion. The upper panel displays the tracks with
spherical accretion while the lower panel shows the result of  channelled accretion on to  hot spots covering 5\% of the surface upon which 75\% of the accretion luminosity is emitted. 
The data points are taken from \citet{2013A&A...550A..21S}, derived from ATCA 18\,GHz observations. The thick red line corresponds to the relationship for ZAMS stars, taken from
\citep{1973AJ.....78..929P}. The model tracks correspond to stars of final mass 
100 (full) , 50 (dotted), 30 (dashed) , 15  (dot-dashed) and 10 M$_\odot$ (3-dot-dashed). \label{lognolyhot}}
\end{figure}

To compare to available data, we plot in Fig.~\ref{lognolyhot} the Lyman photon flux and bolometric luminosity.for a typical hot accretion model, with and without
hot spots.  The observed data are taken from from \citet{2013A&A...550A..21S} and are clearly at variance with this model:  there remains a significant number of data point lying above the tracks. This problem was discussed by  \citet{2013ApJS..208...11L}, \citet{2013MNRAS.tmp.2012U} and \citet{2013A&A...550A..21S}   on comparing data to the expected Lyman flux from ZAMS stars. \citet{2013A&A...550A..21S} speculated that one resolution could be if there was an extra component from the accretion. The lower panel, however, demonstrates that this is  not sufficient in the case  where the star itself has formed through spherical accretion: the bloated protostar is too large to permit a significant release of extreme ultraviolet photons through free-fall on to the surface. This conclusion applies to all accretion models discussed here.

\subsection{Radiative Feedback: cold accretion}

\begin{figure}
\includegraphics[width=8.7cm]{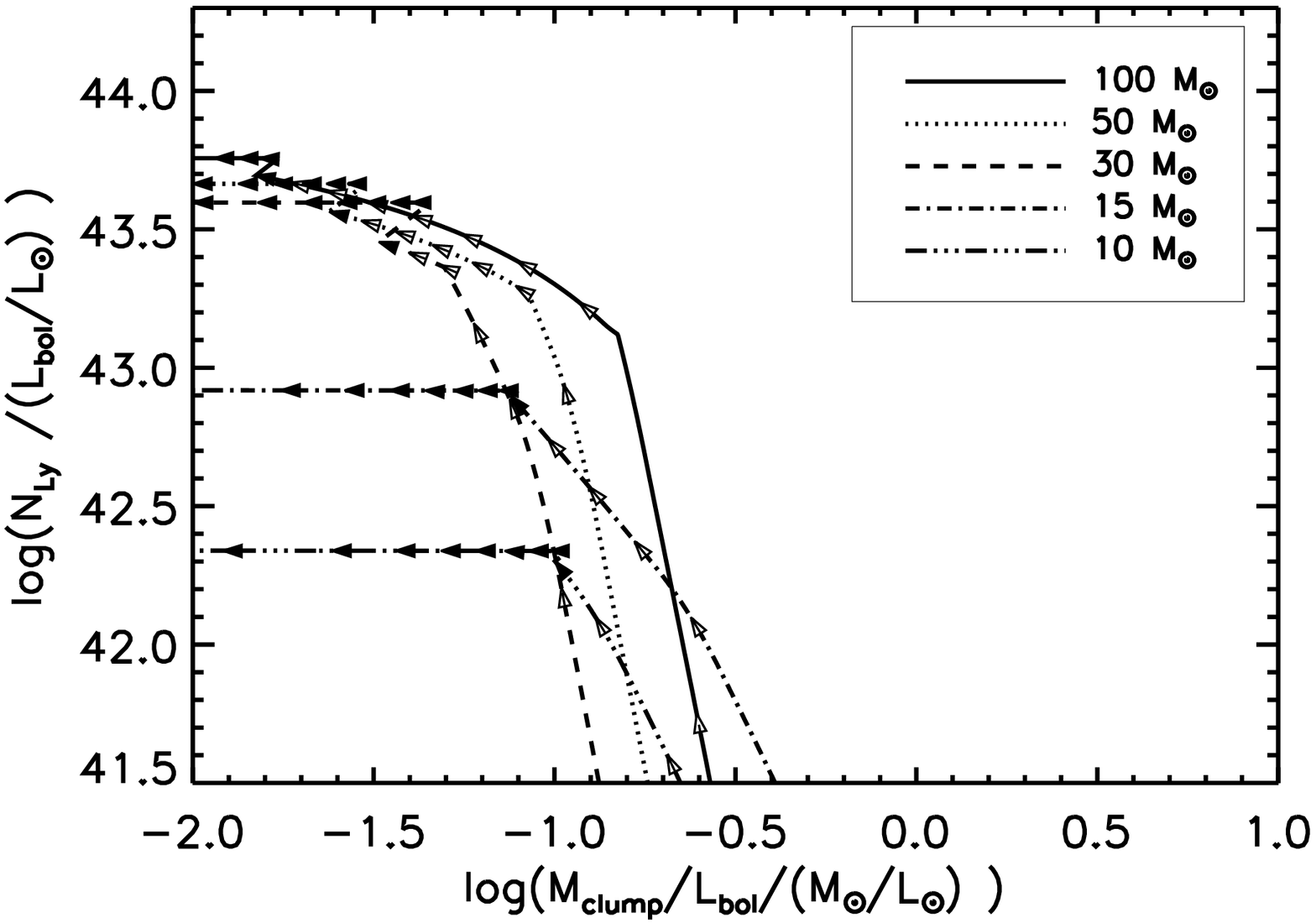}
\includegraphics[width=8.7cm]{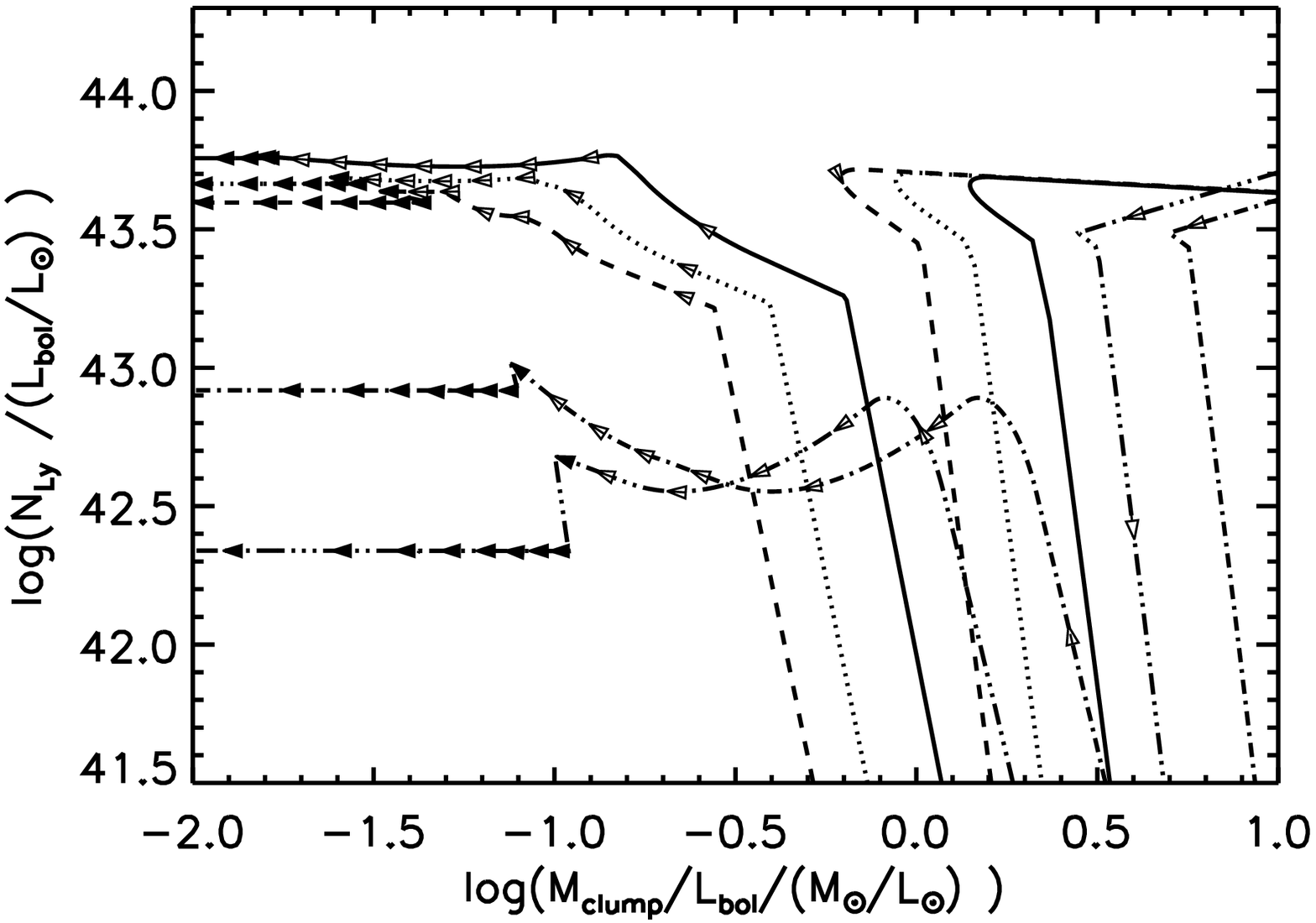}
\caption{Cold accretion. The ratio of Lyman photon flux to bolometric luminosity, $N_{Ly}$/$L_{bol}$, against the evolutionary measure $M_{clump}/L_{bol}$. 
In this diagram, any large distance ambiguity is excluded. The Constant Accretion Model with accretion settling on to the star's surface (upper panel) and with 75\%
being funnelled on to 5\% of the surface at the free-fall speed  (lower panel) are displayed.  \label{lyman-cold}}
\end{figure}

Cold accretion generates a young star with a considerably smaller radius through the early phases. Hence cold accretion does indeed generate more Lyman photons earlier as shown in the top panel of Fig~\ref{lyman-cold} although not greatly different from the hot accretion examples (note the different axial scales).  

We now again suppose that the inner accretion  is guided by the magnetic field to form accretion hot spots on the surface. As shown in the lower panel of Fig~\ref{lyman-cold},
the  behaviour is very different: the accretion luminosity dominates the early UV emission. Moreover, the hotspots are very important Lyman emittors for the stars of mass 
10 -- 20~M$_\odot$ and will  dominate the radio emission during the early phases of star formation. 

\begin{figure}
\includegraphics[width=8.7cm]{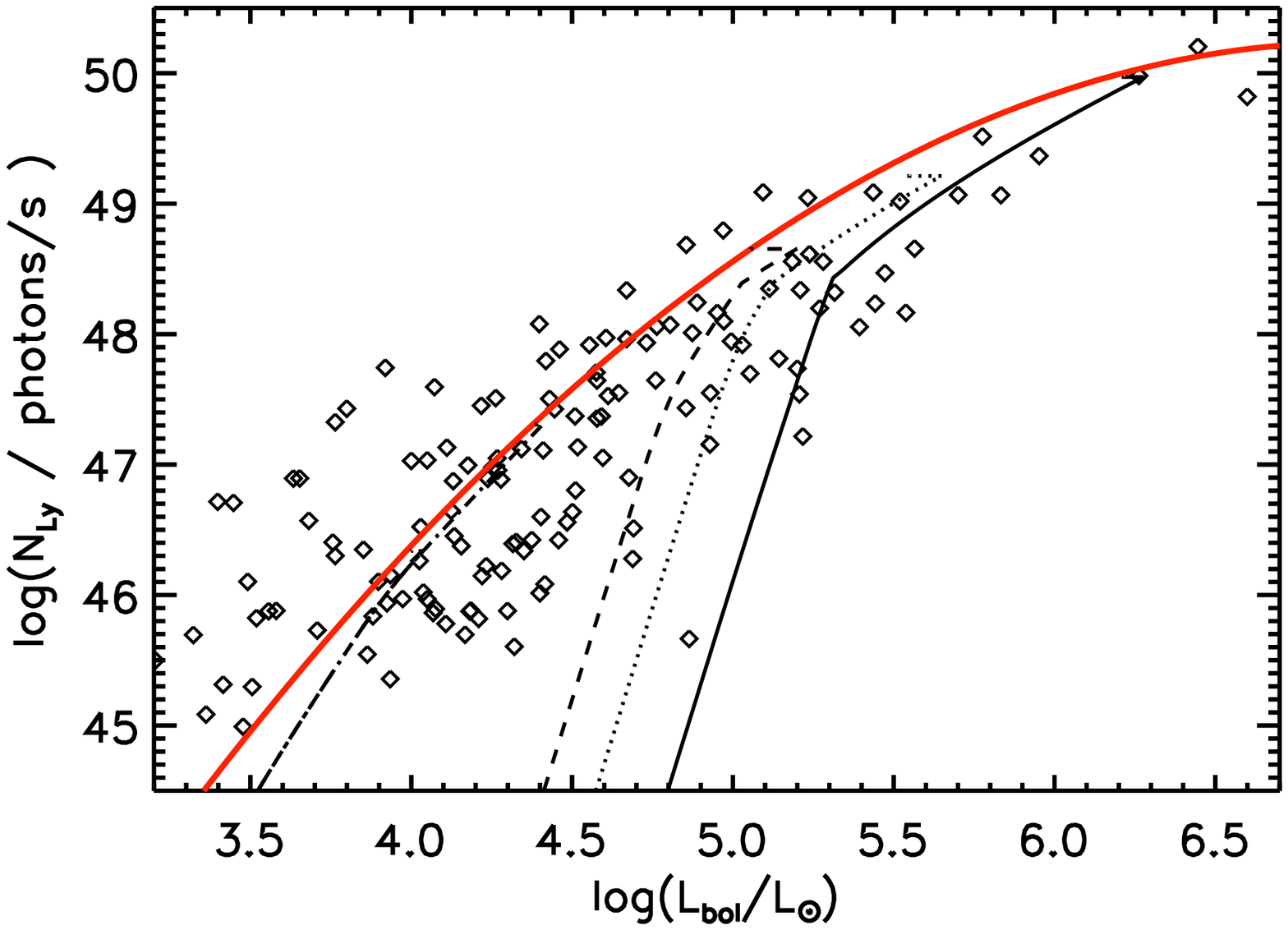}
\includegraphics[width=8.7cm]{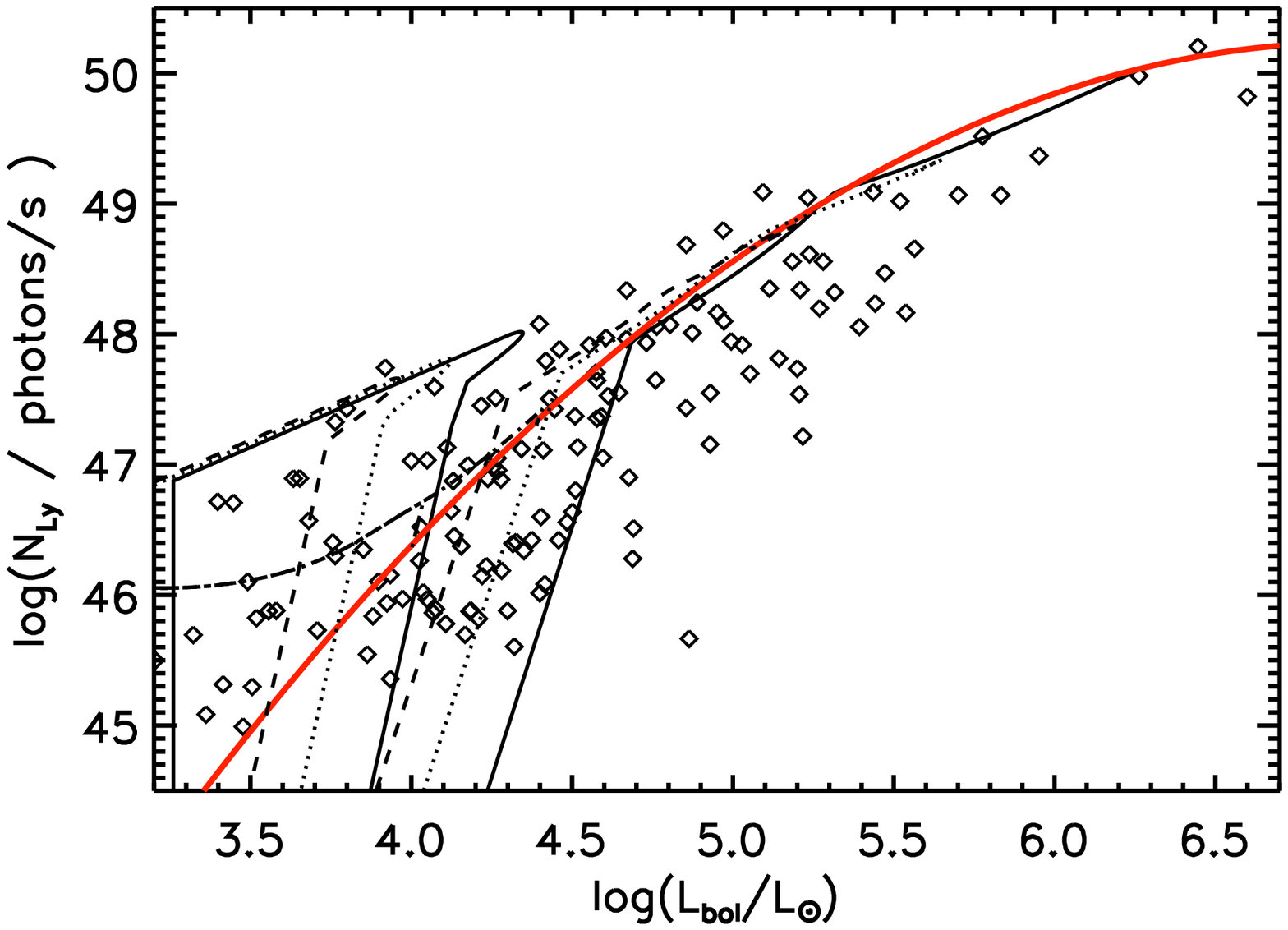}
\caption{The  Lyman photon flux, $N_{Ly}$, against  $L_{bol}$ for the Constant Accretion Model with a compact protostar formed via Cold Accretion. The upper panel displays the tracks with
pure disk accretion while the lower panel shows the result of  channelled accretion on to hot spots covering 5\% of the surface upon which 75\% of the accretion luminosity is emitted. 
The data are taken from \citet{2013A&A...550A..21S}, derived from ATCA 18\,GHz observations. The thick red line corresponds to the relationship for ZAMS stars, taken from
\citet{1973AJ.....78..929P}. The model tracks correspond to stars of final mass 
100 (full) , 50 (dotted), 30 (dashed) , 15  (dot-dashed) and 10 M$_\odot$ (3-dot-dashed). \label{lognolycold}}
\end{figure}

Fig.~\ref{lognolycold}  compares the Constant Accretion Model to the ATCA data.  While cold  accretion does not generate sufficient extreme ultraviolet (top panel)  if a considerable fraction of the accreting material is funnelled on to accretion hotspots, then the data can be very well interpreted (lower panel). This implies that the star has formed through a disk rather than spherical infall. However, at some stage,
the magnetic field becomes sufficiently strong so that material is diverted and funnelled from the inner disk radius to effectively free-fall on to the surface. The star, of course, has
previously formed through the cold accretion and so maintains the relatively small radius.  The enhanced Lyman flux is a result of the high accretion on to a small growing protostar. The subsequent temporary extreme fall in the Lyman flux occurs as the bolometric luminosity falls and the star expands. 

\begin{figure}
\includegraphics[width=8.7cm]{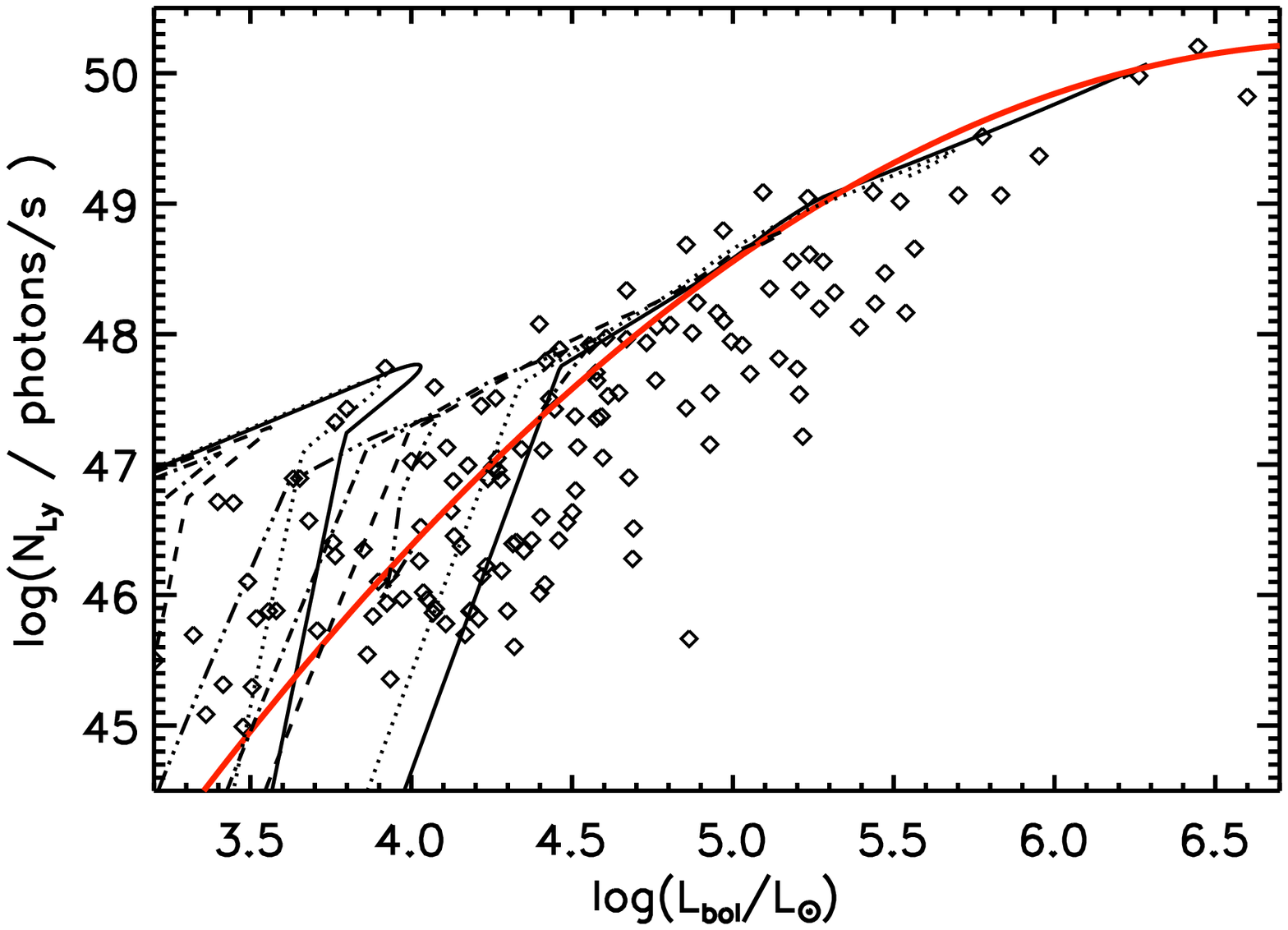}
\includegraphics[width=8.7cm]{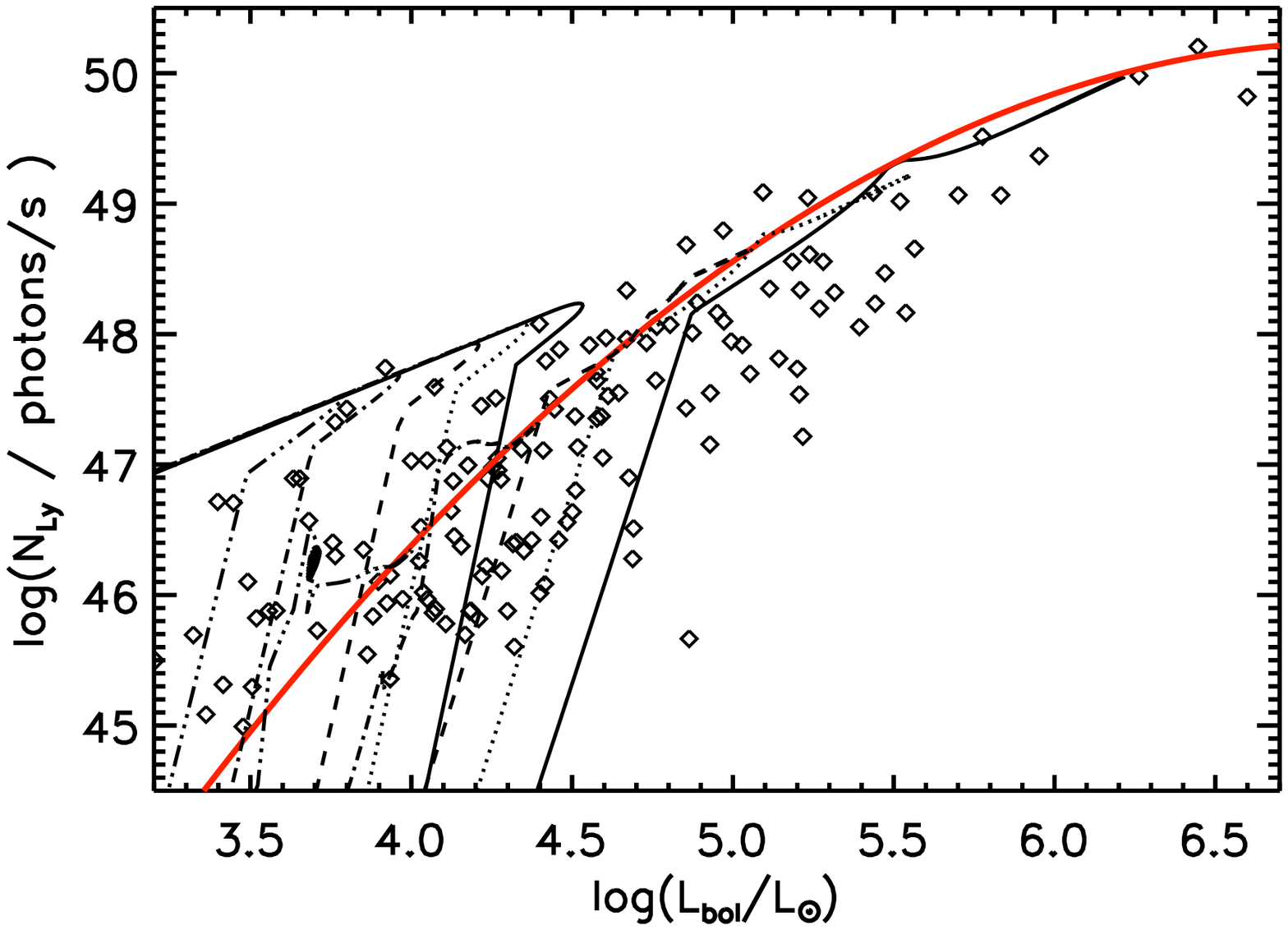}
\caption{The  Lyman photon flux, $N_{Ly}$, against  $L_{bol}$ for the hot spot model combined with with the Accelerated Accretion Model (upper panel) 
and the Power Law Accretion Model (lower panel)  with a compact protostar formed via Cold Accretion. The results of  channelled accretion on to hot spots covering 5\% of the surface upon which 75\% of the accretion luminosity is emitted are displayed.
The data are taken from \citet{2013A&A...550A..21S}, derived from ATCA 18\,GHz observations. The thick red line corresponds to the relationship for ZAMS stars, taken from
\citep{1973AJ.....78..929P}. The model tracks correspond to stars of final mass 
100 (full) , 50 (dotted), 30 (dashed) , 15  (dot-dashed) and 10 M$_\odot$ (3-dot-dashed). \label{lognolycold2}}
\end{figure}

We find that the conclusion that funnelled accretion onto a compact protostar is occurring is independent of the chosen accretion model, as illustrated in Fig.~\ref{lognolycold2} for two extreme accretion types.  Good fits to the ATCA data  require hotspot surface areas of  less than 3\% -- 5\% for
mass fractions of 50\% and 75\% respectively. These ranges are consistent with the fractions deduced from observations of young stars \citep{1998ApJ...509..802C}. However,
it should be noted that while we concentrate on explaining the enigmatic high Lyman flux, most observed sources are either consistent with ZAMS or are underluminous. Some of these sources 
are not consistent with that expected on taking into account the additional low-Lyman flux of the associated stellar cluster  \citep{2013ApJS..208...11L,2013MNRAS.tmp.2012U}.
In the present context, these sources can be the result  of either (1) distributed accretion over the surface and/or (2) single stars in  the early-bloating or late Kelvin-Helmholtz contraction  phases.

 The anomolously high Lyman fluxes only require the formation of intermediate mass stars. High accretion rates are required to explain some data points.  However, as can be seen from Fig.~\ref{lognolycold2}, the protostar need only grow up to the beginning of the bloating phase in order to account for the Lyman flux through hot spot accretion. 
 
\section{Conclusions}

A model for massive stars has been constructed by piecing together models for the protostellar structure, the inflow from a large clump and the 
radiation feedback. The framework requires the accretion rate from the clump to be specified. In this first work we consider a specific subset of possible flows. 
We consider both hot and cold accretion scenarios, identified  as the limiting cases for spherical free-fall and disk accretion, respectively. We assume the fiducial cases
presented by  \citet{2009ApJ...691..823H} in the
`Hot Accretion' scenario and  \citep{2010ApJ...721..478H} for  the  `Cold Accretion' structure 
 but it should be noted that there is considerable uncertainty, depending on the assumed initial interior state and the physics of the radiation feedback
(see also \citet{2013ApJ...772...61K}).  

Strongly variable accretion rates have been investigated by \citet{2012MNRAS.424..457S} as well as  \citet{2013ApJ...772...61K} by utilising
hydrodynamic simulations. In both these works,  the puffy extended nature of the protostars is evident.
Here we only consider smooth evolutions and
 do not consider accretion outbursts or pulsations, or the jet and outflow properties. We also do not consider binary formation, geometry and inclination effects,
or the evolution of the size of the H\,{\small II} region. Hence, this first work sets up the fundamental algorithms and compares results for two recent diagnostic tools. 

Models for the formation of massive stars through accretion can be tested by comparing predictions to a range of observational parameters.
These include the bolometric and extreme ultraviolet luminosities, the envelope and disk mass, and the outflow momentum and energy.
We here determine possible evolutionary tracks on assuming the variation with time of the accretion rate from  a molecular clump on to the star and calculate how the star, envelope and outflow simultaneously evolve. This is achieved by making analytical prescriptions for the components based on current knowledge. We update and extend previous models and confirm previous conclusions that the clump mass must far exceeds the accreted mass, most of it being converted into a surrounding cluster of low-mass objects or dispersed. 
 
We find that Accelerated Accretion is not favoured on the basis of the L$_{bol}$--M$_{clump}$ diagnostic diagram which does not directly provide a test to differentiate the models.
Only a slow accretion model can be distinguished in which the star and clump evolve on the same time scale,  which is not pursued since it seems unlikely given
the contrasting sound-crossing time scales between cores and clumps.  This is mainly because the protostar tends to accrete most of its mass within a short time span in all the other  models. 

Instead, we show that the time spent within each range of bolometric temperature can be closely related to the underlying accretion model. As shown in Table~\ref{tabletime},
Accelerated Accretion generates relatively more sources at low bolometric temperatures. Sets of far-infrared  Herschel data covering the temperature range from 20\,K -- 70\,K
should provide some insight. However, modelling and observations of the bolometric temperature both remain problematic especially at the low temperatures which can be dominated by unbound non-stellar and pre stellar objects in addition to AGB stars \citep{2013A&A...549A.130V}.  However, we find a solution which fits the data in which the initial clump mass is four to five times larger than that necessary to generate the associated star cluster corresponding to the mass of the most massive star. We thus generate revised evolutionary tracks which are consistent with statistics for the bolometric temperature. In these revised models, the star remains deeply embedded throughout its formation and the bolometric temperature distribution is no longer a sensitive diagnostic to differentiate between accretion models.

Accretion models may be better tested through a complete sample of hotter  
protostars with temperatures in the range 50\,K to 100\,K. In addition, as with their low-mass   counterparts, large periodic accretion variations could dominate the statistics. 

The Accelerated Accretion Model has been advanced in the literature because we expect the gravitational sphere of  influence of a central object to grow as its mass grows.
Here, however, with the large accretion rates often assumed, we take an inner envelope to already exist with sufficient mass to meet the later needs of the star and outflow.
In this case, the accretion rate depends on how fast mass can flow in from the envelope, through the disk, rather than how massive the central protostar has become.

Finally, we have investigated Lyman fluxes as deduced observationally from radio fluxes. Observationally, it has been shown that objects of luminosity $\sim$ 10,000~L$_\odot$ can possess very high Lyman fluxes, inconsistent with their expected stellar temperatures \citep{2013A&A...550A..21S}. We have shown here that the problem is resolvable if the protostar is relatively compact, formed through cold accretion via a disk. However, the present accretion must
involve a funnelling free-fall  mechanism onto a fraction of the stellar surface estimated to be less than 10\%. The mechanism for this remains unknown but if the evidence that massive protostars are configured in the same way as low-mass stars continues to grow, then accretion via magnetic flux tubes and jets driven by magneto-centrifugal processes are conceivable.

The above explanation of the excess Lyman photon flux requires evolution under the Cold Accretion scenario. Hot Accretion falls far short even with the inclusion of hot-spots, as demonstrated in Fig.~\ref{lognolyhot}. Cold accretion was defined as accretion on to the photosphere with no back-heating, so that the accreting material has the same entropy as that of the photosphere \citep{2010ApJ...721..478H}. This is a limiting case which may be difficult to realise. It could be generally expected that the rapid mass accretion should be somewhat hot because a fraction of the entropy should be advected into the stellar interior \citep{2011arXiv1106.3343H}. Taking larger protostellar radii will reduce the Lyman flux;  taking radii 50\% higher than that predicted for cold accretion, however, does not significantly alter the qualitative of fit to the ATCA data while doubling the radius has a considerable effect. Accurate results will require the implementation of a full
stellar evolution code.  

The major objective here has been to construct a consistent model which links the components. In the following works, we will investigate the consequences of mass
outflows, accretion variations, maser production, thermal radio jets and H~{\small II} regions with the purpose of determining how their evolutions are coordinated.  

\section{Acknowledgements}

I wish to thank  Riccardo Cesaroni, Davide Elia, Sergio Molinari and Alvaro Sanchez-Monge for their encouragement and comment.

\end{document}